\documentclass[aps,twocolumn,showpacs,preprintnumbers,amsmath,amssymb]{revtex4}

\usepackage{graphicx}
\usepackage{epstopdf}
\usepackage{bm}
\usepackage[dvipsnames]{color}

\begin{document}

\def \brho{{\hbox{\boldmath $\rho$}}}
\def \beps{{\hbox{\boldmath $\varepsilon$}}}
\def \bdelta{{\hbox{\boldmath $\delta$}}}

\title{Optical absorption and conductivity in quasi-two-dimensional crystals from first principles: Application to graphene}
\author{Dino Novko$^{1}$}
\author{Marijan \v Sunji\'c$^{1,2}$}
\author{Vito Despoja$^{1,2,3}$}
\affiliation{$^1$Donostia International Physics Center (DIPC), P. Manuel de Lardizabal, 20018 San Sebastian, Basque Country, Spain}
\affiliation{$^2$Department of Physics, University of Zagreb, Bijeni\v{c}ka 32, HR-10000 Zagreb, Croatia}
\affiliation{$^3$Universidad del Pais Vasco, Centro de Fisica de Materiales CSIC-UPV/EHU-MPC, Av. Tolosa 72, E-20018 San Sebastian, Spain}

\begin{abstract}
This paper gives a theoretical formulation of the electromagnetic response of the quasi-two-dimensional (Q2D) crystals suitable for investigation of optical activity and polariton modes. The response to external electromagnetic field is described by current-current response tensor $\Pi_{\mu\nu}$ calculated by solving the Dyson equation in the random phase approximation (RPA), where 
current-current interaction is mediated by the photon propagator $D_{\mu\nu}$. The irreducible current-current response tensor $\Pi^0_{\mu\nu}$ is calculated from the {\em ab initio} Kohn-Sham (KS) orbitals. The accuracy of $\Pi^0_{\mu\nu}$ is tested in the long wavelength limit where it gives correct Drude dielectric function and conductivity. The theory is applied to the calculation of optical absorption and conductivity in pristine and doped single layer graphene and successfully compared with previous calculations and measurements. 
\end{abstract}

\maketitle

\section{Introduction}
Understanding the interaction between light and electrons in a crystal has always been an attractive topic, and its extensive study led to  the realization of many devices, such as lasers, semiconducting solar cells or LED diodes. More recently it led to new discoveries, such as sub-wavelength light transmission \cite{Ebb}, waveguiding \cite{waveguid}, hybrid solid state/organic solar cells \cite{solarcell}, etc. There are many theoretical models which successfully deal with these phenomena, mostly based on solving Maxwell's equations at the boundaries of the crystals of different shapes \cite{Vidal} and different dielectric properties, calculated at different levels of approximations, e.g. within the Drude dielectric model \cite{OptT,Pol2}, or from first principles and beyond the random phase approximation (RPA) \cite{Rubio}. 

However, what is still missing is a theoretical approach where the interaction between light and crystal electrons would be calculated fully microscopically, so that the electronic structure is calculated using {\em ab initio} methods (usually in the simplified tight-binding or subband models \cite{tb1,tb2,tb3,tb4,sb}), the polarizability of the system is described by the current-current response tensor (usually by the density-density response function \cite{comment}) and where the electron-electron interaction is mediated by photons (usually described by instantaneous Coulomb interaction). Inclusion of these effects could be crucial if one wanted to explore new optically active (radiative) modes or self-sustainable electromagnetic modes (polaritons) in crystals.         

The aim of this paper is to give a theoretical formulation of the interaction between electromagnetic field and electronic excitations in quasi-two-dimensional (Q2D) crystals (consisting of one or few atomic layers), suitable for investigation of optically active electronic modes and polaritons. This formulation was partially developed in Ref. \cite{Pol} where it was applied to investigate electromagnetic modes in the jellium metallic slab. Here this formulation is extended to the case where the ground state electronic structure is calculated from first principles. To test the theory we calculate the optical absorption and conductivity in the self standing graphene monolayer and compare our results with the recent experimental and theoretical works \cite{Louie,SSC,2.3posto1,2.3posto2,No1,No2}. 

Optical properties of graphene have already been extensively investigated, both from the experimental and theoretical viewpoints. Apart from the above mentioned works, the information about optical activity of $\pi$  or $\pi+\sigma$ plasmons or single particle excitations was also extracted from electron energy loss (EELS) experiments and corresponding calculations \cite{RubioGraph,EELS1,EELS2,EELS3,EELS4,Gr2013,Dino}. In some cases the dispersion relations of 2D polaritons and conductivity in doped graphene are calculated using RPA density-density response function \cite{Pol1,sarma1,Gr2013,Silkin} or using additionally a phenomenological relaxation-time approximation to account for the damping effects \cite{Pol2,Kupcic}. In Ref. \cite{Kupcic,Kupcic1} optical properties and conductivity in graphene are investigated at a high level of accuracy, beyond RPA, however the orbital and band structure are described within the tight binding approximation (TBA). On the other hand, in Ref. \cite{Louie} optical properties of graphene are calculated from first principles including quasiparticle corrections and solving Bethe-Salpeter equation (so-called GW-BSE scheme) for the polarizability tensor. This GW-BSE scheme includes excitonic effects properly, resulting in a nice agreement of ultraviolet (UV) active $\pi\rightarrow\pi^*$ peak with the experimental measurements, while our RPA theory underestimates this experimental value. Nevertheless, our theory provides very good agreement with the experiment in the infrared (IR) and visible regions and is capable to calculate the optically inactive (evanescent) modes such as surface plasmon polaritons (SPP).

It is worth mentioning, due to recent interest \cite{Nazarov, Nazarov1, Politano15, Sohier}, that methodology presented in this paper can be adopted to obtain the charge-charge response function $\chi(\mathbf{Q},\omega)$ and the dielectric function $\varepsilon(\mathbf{Q},\omega)$ of Q2D material by connecting the former with current-current correlation function $\Pi_{\mu\nu} (\mathbf{Q},\omega)$ using the gauge invariance and the conservation of local charge density \cite{Schrieffer,Kupcic}. Numerically this can be very convenient for obtaining $\varepsilon(\mathbf{Q},\omega)$ within RPA response, because the $\mathbf{Q}^2$ divergence of the bare Coulomb interaction is automatically cancelled due to $\mathbf{Q}$ prefactor in current vertex function and the special care for $\mathbf{Q}=0$ case \cite{Olsen} is not needed.

In Sec. \ref{Formu} we first present the general formulation of the problem, description of the system and the derivation of the Dyson equation for the screened current-current response tensor $\Pi_{\mu\nu}$. In Sec. \ref{Calpi} we formulate the Dyson equation for a specific geometry of the system in terms of Kohn-Sham (KS) electronic wave functions. In Sec. \ref{Abscond} the expressions for optical absorption and conductivity are given in terms of the tensor $\Pi_{\mu\nu}$.  During {\em ab initio} calculation of nonlocal paramagnetic and local diamagnetic terms in $\Pi_{\mu\nu}$ certain numerical problems arise which are discussed in Sec. \ref{HAlter} and resolved using an alternative expression for the current-current response tensor. In Sec. \ref{dugi} we prove that alternative expression of $\Pi^0_{\mu\nu}$ in the long wavelength limit leads to the Drude dielectric function and conductivity. In Sec. \ref{Results} we apply the developed methodology to calculate the optical absorption spectrum and conductivity in doped and pristine graphene and compare it with recent experimental and theoretical results. Sec. \ref{Comp} gives details of the computational procedure. We use density functional theory (DFT) ground state calculation to get the crystal KS orbitals $\phi_{n{\bf K}}$ and band structure $E_{n{\bf K}}$. For the ground state calculation we use PWSCF method which means that our crystal should be 3D periodic. Here the superlattice consists of periodically repeated supercells containing several atomic layers (crystal slab). We also explain how to avoid the effect of interaction with the neighboring supercells. In the second stage of the calculation we solve the Dyson equation for current-current response tensor $\Pi_{\mu\nu}$ where irreducible current-current response tensor $\Pi^0_{\mu\nu}$ enters and electromagnetic interaction is mediated by the photonic propagator $D_{\mu\nu}$ \cite{Pol}. Here we restrict our calculations to RPA where irreducible current-current response tensor $\Pi^0_{\mu\nu}$ can be obtained directly from the crystal KS states. A problem arises from the fact that we want to investigate optical properties of single Q2D crystal slab, while electronic structure is calculated for the entire 3D superlattice. This problem can not be solved simply by increasing the vertical separation between slabs because now each of them radiates electromagnetic field which spreads across the entire space and interaction between slabs is unavoidable. We solve this by cutting off the vertical range of the photon propagator $D_{\mu\nu}$ and allowing propagation only within one Q2D crystal slab. This procedure allows smaller vertical distances between adjacent slabs (enough to avoid the overlap between their charge densities) while still cancelling the interaction between them. Similar method is successfully utilized for calculation of EELS spectrum in graphene \cite{Gr2013,Dino}. In Sec. \ref{lightabs} we present and discuss in detail results for optical absorption spectra in pristine and doped graphene. In Sec. \ref{Conclu} we present the conclusion.

\section{Formulation of the problem}
\subsection{Derivation of the current-current response tensor} 
\label{Formu}
In this section we will first derive the Dyson equation for the screened current-current response tensor in the Q2D crystal consisting of one or few atomic layers. We consider independent electrons which move in a local (DFT) crystal potential and interact with the electromagnetic field described by the vector potential operator ${\bf A}({\bf r},t)$. Then the Hamiltonian of the system can be written as 

\begin{equation}
H=H_0^{e}+H_0^{\mathrm{EM}}+V^{\mathrm{in}}
\label{Ham}
\end{equation}
where 

\begin{equation}
H_0^{\mathrm{e}}=\sum_{{\bf K},n}E_{n{\bf K}}c^{\dag}_{n{\bf K}}c_{n{\bf K}}
\label{ineham}
\end{equation}
describes non-interacting electrons in some local potential. Here $c^{\dag}_{n{\bf K}}$ is the creation operator of an electron in the Bloch state $\left\{n,{\bf K}\right\}$, with the wave function $\phi_{n{\bf K}}\left({\bf r}\right)$ and energy $E_{n{\bf K}}$, where $n$ is the band index and ${\bf K}$ is the electron wavevector in the plane parallel to crystal layers. $H_0^{\mathrm{EM}}$ is the Hamiltonian of free electromagnetic field. In the $\Phi=0$ gauge the interacting Hamiltonian   

\begin{equation}
V^{\mathrm{in}}=V^{\mathrm{p}}+V^{\mathrm{d}}
\label{internalInt}
\end{equation}
consists of the paramagnetic part 

\begin{equation}
V^{\mathrm{p}}=-\frac{1}{c}\int d{\bf r}\ {\bf j}({\bf r}) {\bf A}({\bf r})   
\label{Paraint}
\end{equation}
and the diamagnetic part 

\begin{equation}
V^{\mathrm{d}}=\frac{e^2}{2mc^2}\int d{\bf r}\ \rho({\bf r}){\bf A}^2({\bf r}).
\label{Diaint}
\end{equation}
Here the electron current operator is  

\begin{equation}
{\bf j}({\bf r})=\frac{e\hbar}{2im}
\left\{
\psi^{\dag}\left({\bf r}\right)\nabla\psi\left({\bf r}\right)
-
\left[\nabla\psi^{\dag}\left({\bf r}\right)\right]\psi\left({\bf r}\right)\right\},
\label{parcurr}
\end{equation}
electron density operator is 

\begin{equation}
\rho({\bf r})=\psi^{\dag}\left({\bf r}\right)\psi\left({\bf r}\right), 
\label{densop}
\end{equation}
and electron field operators are  

\begin{equation}
\psi\left({\bf r}\right)=\sum_{n,{\bf K}}\phi_{n{\bf K}}({\bf r})c_{n{\bf K}}. 
\label{oppo}
\end{equation}

\begin{figure}
\centering
\includegraphics[width=\columnwidth]{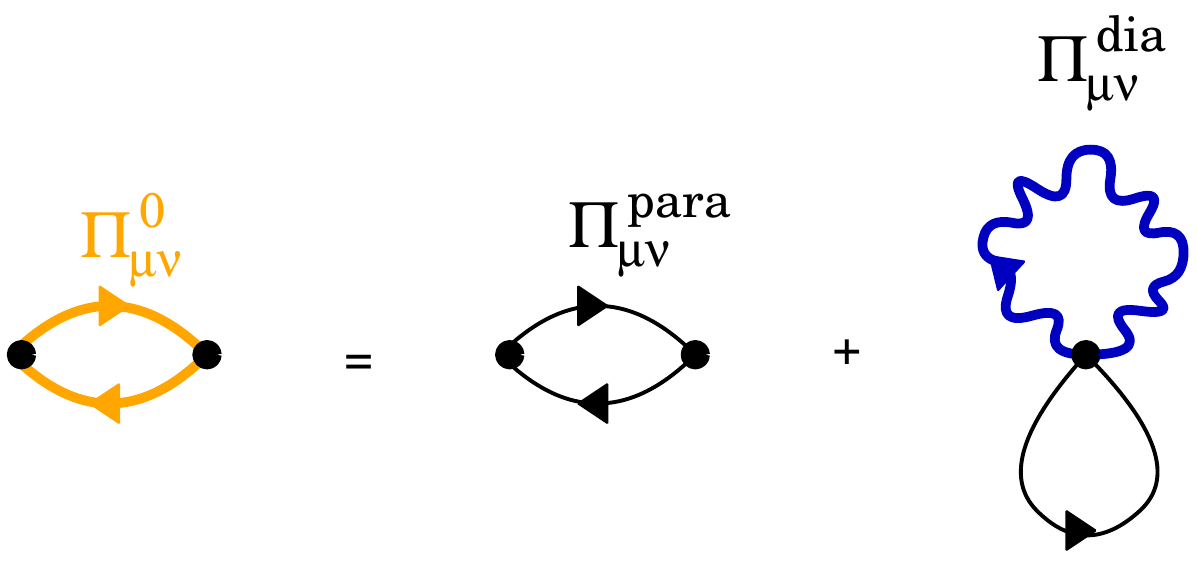}
\caption{(Color online) Diagrammatic representation of the non-interacting current-current response tensor and its paramagnetic and diamagnetic terms (\ref{current-current}).}
\label{Fig1a}
\end{figure}

\begin{figure}
\centering
\includegraphics[width=\columnwidth]{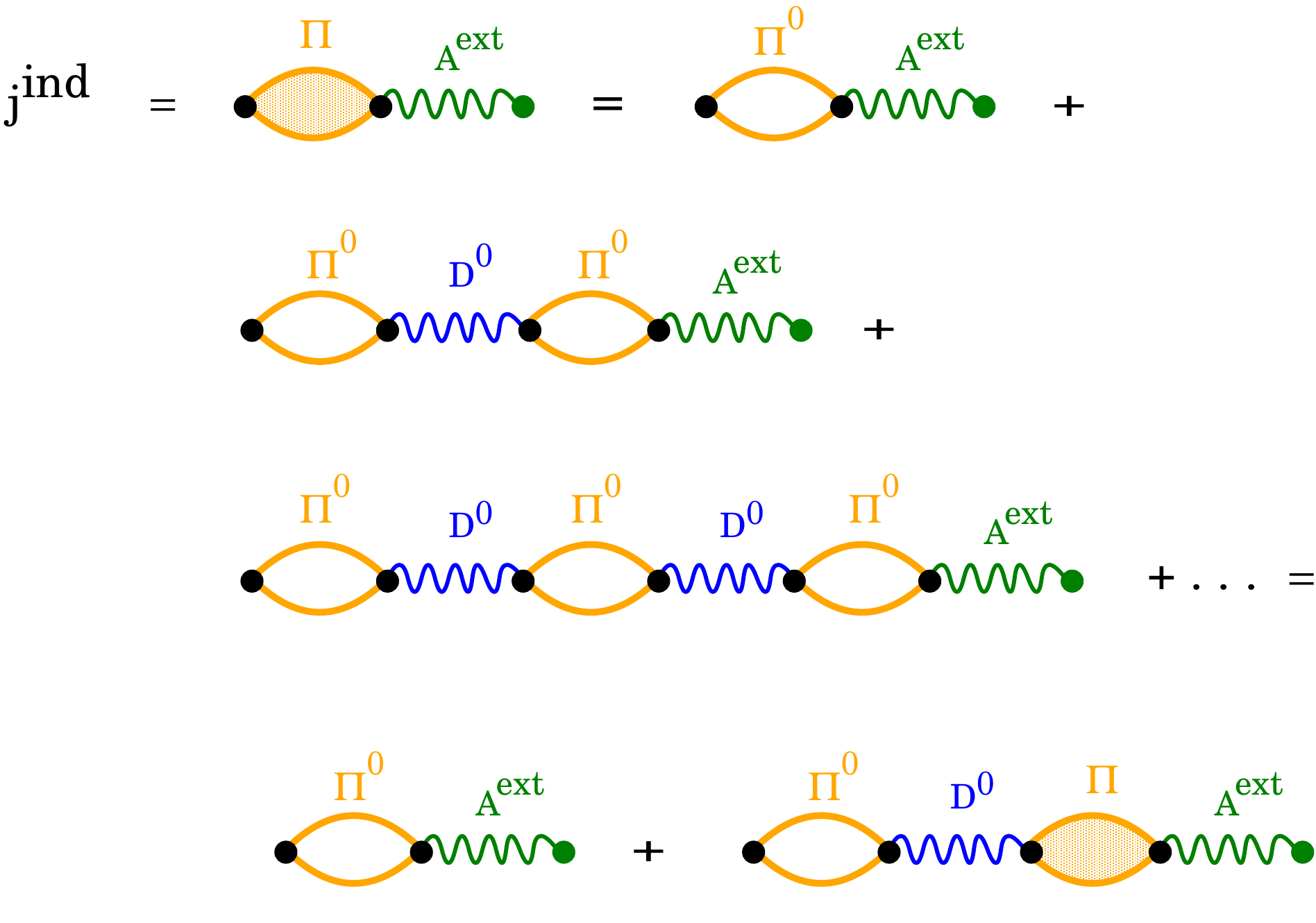}
\caption{(Color online) Diagrammatic expansion of the induced current within RPA (\ref{zaj}). $\Pi^0$ represents non-interacting current-current response tensor (\ref{current-current}), $\Pi$ interacting current-current response tensor within RPA (\ref{Dyseq}), and $D^0$ free photon propagator (\ref{Freephotonpropagator}).}
\label{Fig1}
\end{figure}

The key quantity that will give the physical properties (e.g. optical properties) of the system of electrons interacting with the 'internal' electromagnetic field (electron-electron interaction mediated by photons) will be the current-current response tensor $\Pi_{\mu\nu}({\bf r},{\bf r}',t,t')$, which in the 0-th order of the perturbation expansion over interaction $V^{\mathrm{in}}$ contains two terms: 
\begin{equation}
\Pi^0_{\mu\nu}=\Pi^{\mathrm{para}}_{\mu\nu}+\Pi^{\mathrm{dia}}_{\mu\nu}.
\label{current-current}
\end{equation}       
Here the paramagnetic term is:  

\begin{eqnarray}
\Pi^{\mathrm{para}}_{\mu\nu}({\bf r},{\bf r}',t,t')=\hspace{5cm}
\nonumber\\
\label{Self1}\\
\frac{i}{\hbar c}\theta(t-t')
\left\langle\Psi^0_{e}\left|\left[j_{\mu}({\bf r},t),
j_{\nu}({\bf r}',t')\right]_-\right|\Psi^0_{e}\right\rangle
\nonumber
\end{eqnarray}
and $\left|\Psi^0_{e}\right\rangle$ is the ground state of the Hamiltonian $H^{e}_0$. The diamagnetic term is

\begin{equation}
\Pi^{\mathrm{dia}}_{\mu\nu}\left({\bf r},{\bf r}',t,t'\right)=
-\frac{e^2}{mc}
n({\bf r})\delta(t-t')
\delta({\bf r}-{\bf r}')\delta_{\mu\nu} 
\label{Self2}
\end{equation}
where $n({\bf r})=\left\langle\Psi^0_{e}\left|\rho({\bf r})\right|\Psi^0_{e}\right\rangle$ represents the ground state electron density. In the lowest order these two terms in the expansion of the irreducible current-current response tensor are represented diagrammatically in Fig. \ref{Fig1a}. Next step is to provide perturbation expansion for the tensor $\mathbf{\Pi}$ which now includes the higher order terms in the interaction $V^{\mathrm{in}}$. If we restrict our consideration within RPA the perturbation expansion of the current-current response tensor (\ref{current-current}) becomes:

\begin{eqnarray}
{\bf \Pi}={\bf \Pi}_0+{\bf \Pi}_0\otimes{\bf D}_0\otimes{\bf\Pi}_0+\hspace{3cm}
\nonumber\\
\label{Dysexp}\\
{\bf \Pi}_0\otimes{\bf D}_0\otimes{\bf\Pi}_0\otimes{\bf D}_0\otimes{\bf\Pi}_0+\ ...\ \hspace{1cm}
\nonumber
\end{eqnarray}
where the symbol $\otimes$  denotes the convolution or integration over spatial and time variables $({\bf r}, t)$ in addition to the matrix multiplication over indices $\mu=x,y,z$. Also for clarity we omit to write spatial and time variables. From the expansion (\ref{Dysexp}) it is obvious that the 'screened' current-current response tensor can be calculated by solving the Dyson equation 

\begin{equation}
{\bf \Pi}={\bf \Pi}_0+{\bf \Pi}_0\otimes{\bf D}_0\otimes{\bf\Pi},\hspace{3cm}
\label{Dyseq}
\end{equation}
where the only inputs are the non-interacting current-current response tensor ${\bf\Pi}_0$ given by (\ref{current-current}) and the retarded free photon propagator given by  

\begin{eqnarray}
D^0_{\mu\nu}\left({\bf r},{\bf r}',t,t'\right)=\hspace{5cm} 
\nonumber\\
\label{Freephotonpropagator}
\\
\frac{i}{\hbar c}\theta\left(t-t'\right)
\left\langle\Psi^0_{\mathrm{EM}}\left|\left[A^0_{\mu}\left({\bf r},t\right),A^0_{\nu}\left({\bf r}',t'\right)\right]_-\right|\Psi^0_{\mathrm{EM}}\right\rangle,
\nonumber
\end{eqnarray}
where $\Psi^0_{\mathrm{EM}}$ is the photon vacuum (ground state of $H^{\mathrm{EM}}_0$), and the operator $A_{\mu}^0$ is defined as:

\begin{equation}
A^0_{\mu}\left({\bf r},t\right)=e^{iH^{\mathrm{EM}}_0 t}A_{\mu}\left({\bf r}\right)e^{-iH^{\mathrm{EM}}_0 t}.
\end{equation}
The perturbation expansion of the current-current response tensor ${\bf \Pi}$ is diagrammatically presented in Fig. \ref{Fig1}, where the green wavy line represents the external field ${\bf A}^{\mathrm{ext}}$, that induces current fluctuations in the crystal. Blue wavy lines represent the propagator of electromagnetic field ${\bf D}^{\mathrm{0}}$, that mediates electromagnetic interaction within the crystal.
It should be mentioned here that this expansion actually goes beyond RPA. Namely, current-current response function (\ref{current-current}) can be calculated by means of single-particle Green's functions. On the other hand electron-electron interaction is partially (depending on the DFT approximation used) included in the Bloch states used to construct the single-particle Green's function. Therefore the electron-electron interaction is included already in the a lowest order of the expansion (\ref{Dysexp}) in the form of self-energy corrections to the irreducible polarizability $\Pi^0$.

The response tensor $\Pi_{\mu\nu}$ contains information about spectroscopic properties of the system (single-particle or collective electromagnetic modes in the system) or information about the response to the external electromagnetic field. Suppose that the crystal is exposed to the external (classical) electromagnetic field described by the vector potential ${\bf A}^{\mathrm{ext}}({\bf r},t)$. In this case the total Hamiltonian (\ref{Ham}) gets an additional term  $V^{\mathrm{ext}}$ which has the form analogous to (\ref{internalInt}--\ref{Diaint}) except that now  ${\bf A}$ should be replaced by ${\bf A}^{\mathrm{ext}}$. If $V^{\mathrm{ext}}$ is a small perturbation it is sufficient to keep only the term linear in ${\bf A}^{\mathrm{ext}}$. In this case the current induced by the external field becomes 

\begin{equation}
j_\mu^{\mathrm{ind}}({\bf r},t)=\sum_\nu\int\ d{\bf r}_1,dt_1 \Pi_{\mu\nu}({\bf r},{\bf r}_1,t-t_1)A_\nu^{\mathrm{ext}}({\bf r}_1,t_1),
\label{zaj}
\end{equation} 
as shown schematically in Fig. \ref{Fig1}. The induced charge density fluctuations is similarly given by 

\begin{equation}
\rho^{\mathrm{ind}}({\bf r},t)=\sum_\mu\int\ d{\bf r}_1,dt_1 \Pi_{0\mu}({\bf r},{\bf r}_1,t-t_1)A_\mu^{\mathrm{ext}}({\bf r}_1,t_1)
\label{roire}
\end{equation} 
where we introduce the density-current response function which is (in the lowest order in expansion over $V^{\mathrm{in}}$) given by 

\begin{eqnarray}
\Pi_{0\nu}^0({\bf r},{\bf r}',t,t')=\hspace{5cm}
\nonumber\\
\label{charcurt}\\
\frac{i}{\hbar c}\theta(t-t')
\left\langle\Psi^0_{e}\left|\left[\rho({\bf r},t),
j^{\mathrm{para}}_{\nu}({\bf r}',t')\right]_-\right|\Psi^0_{e}\right\rangle. 
\nonumber
\end{eqnarray}
Induced current and density fluctuations are connected by the continuity equation

\begin{equation}
\nabla{\bf j}^{\mathrm{ind}}+\frac{\partial}{\partial t}\rho^{\mathrm{ind}}=0. 
\label{nabcon}
\end{equation}

\subsection{Calculation of the screened current-current response tensor}
\label{Calpi}
\begin{figure}[b]
\centering
\includegraphics[width=\columnwidth]{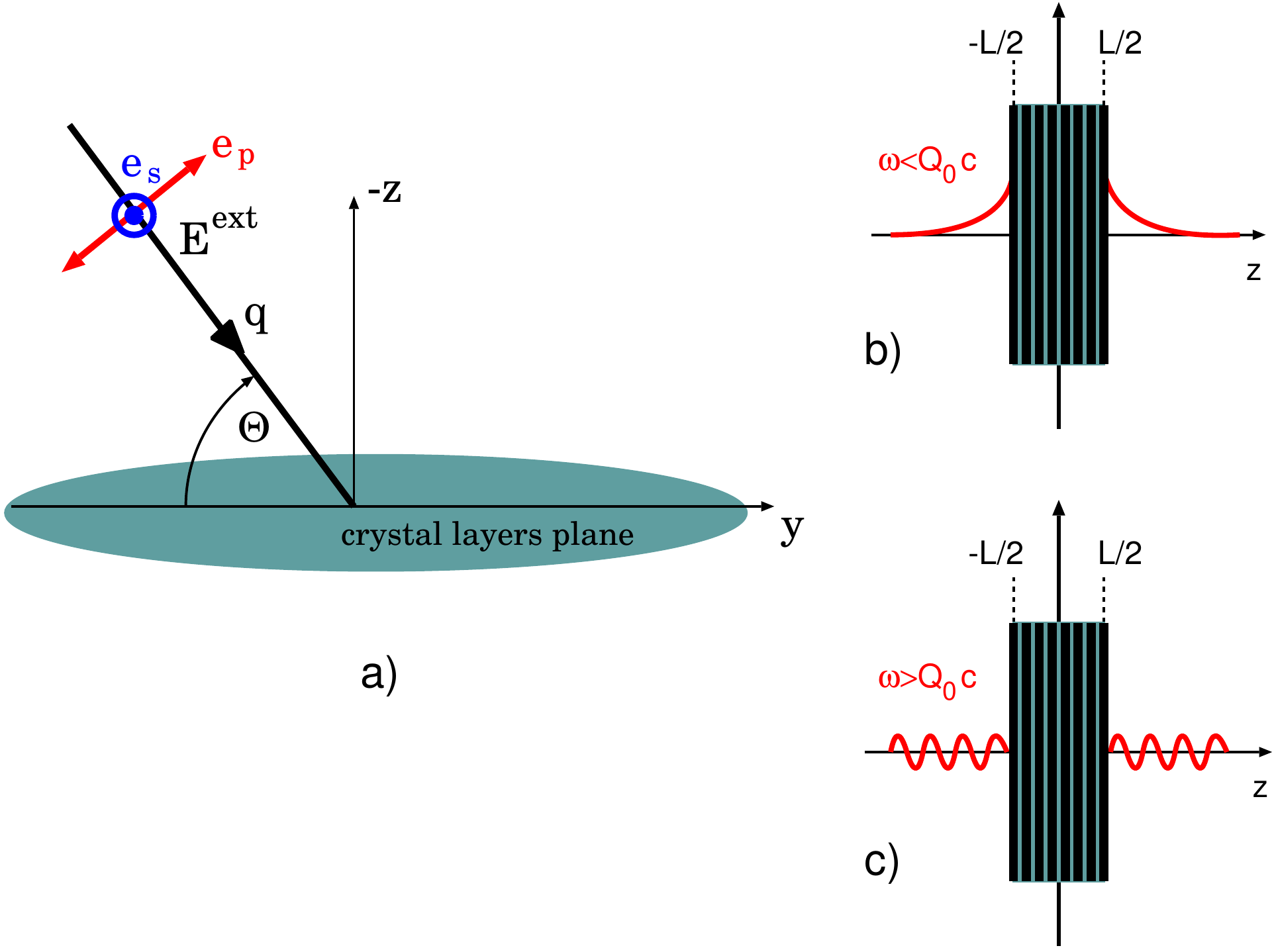}
\caption{(Color online) Geometry of the system. (a) The orientation of the crystal layers and incident electromagnetic field. (b) Evanescent character of the incident electromagnetic field for $\omega<\mathrm{Qc}$. (c) Radiative character of the incident electromagnetic field for $\omega>\mathrm{Qc}$. Layered crystal electronic density is restricted in the region $-\frac{L}{2}<z<\frac{L}{2}$.}
\label{Fig2}
\end{figure}

In this calculation we shall first exploit the symmetry of the system which leads to the conservation of the $\mathbf{Q}$ vector parallel to the surface. Symmetry of the system also 
enables division to $\mathbf{s}$ and $\mathbf{p}$ polarizations. Suppose that the layered crystal ground state electronic density is in the perpendicular $z$ direction restricted to the region $-L/2<z<L/2$, as sketched in Figs. \ref{Fig2}(b) and \ref{Fig2}(c). Crystal periodicity is broken in the $z$ direction but remains in $xy$ plane, so it is appropriate to perform the Fourier transform of the Dyson equation (\ref{Dyseq}) in the $xy$ plane:     

\begin{widetext}

\begin{eqnarray}
{\bf\Pi}_{{\bf G}_\parallel,{\bf G}'_\parallel}({\bf Q},\omega,z,z')=
{\bf\Pi}^0_{{\bf G}_\parallel,{\bf G}'_\parallel}({\bf Q},\omega,z,z')+\hspace{10cm}
\nonumber\\
\label{DysFur}\\
\hspace{4cm}\sum_{{\bf G}_{\parallel 1},{\bf G}_{\parallel 2}}\int^{\frac{L}{2}}_{-\frac{L}{2}}dz_1dz_2\ 
{\bf\Pi}^0_{{\bf G}_\parallel,{\bf G}_{\parallel 1}}({\bf Q},\omega,z,z_1)
{\bf D}^0_{{\bf G}_{\parallel 1},{\bf G}_{\parallel 2}}({\bf Q},\omega,z_1,z_2)
{\bf\Pi}_{{\bf G}_{\parallel 2},{\bf G}'_\parallel}({\bf Q},\omega,z_2,z'),
\nonumber
\end{eqnarray}
where ${\bf G}_\parallel=(G_x,G_y)$ are 2D reciprocal lattice vectors. In (\ref{DysFur}) we also simultaneously performed the Fourier transform in $\omega$-space. The Fourier transform of the free-photon propagator (\ref{Freephotonpropagator}) is given by \cite{Pol}  

\begin{equation}
{\bf D}^0_{{\bf G}_{\parallel},{\bf G}'_\parallel}({\bf Q},\omega,z,z')
={\bf D}^0({\bf Q}+{\bf G}_{\parallel},\omega,z,z')\delta_{{\bf G}_{\parallel},{\bf G}'_\parallel}
\label{photpar}
\end{equation}
where \cite{Green}

\begin{equation}
{\bf D}^0({\bf Q},\omega,z,z')=-\frac{4\pi c}{\omega^2}\delta(z-z')
{\bf z}\cdot{\bf z}  
+\frac{2\pi i}{c\beta}
\left\{
{\bf e}_{s}\cdot{\bf e}_{s}+{\bf e}_{p}\cdot{\bf e}_{p}
\right\}e^{i\beta\left|z-z'\right|}.
\label{photQ}
\end{equation} 
Here the unit vectors are adapted to the geometry of the system such that ${\bf e}_{s}={\bf Q}_{0}\times{\bf z}$ and ${\bf e}_{p}=\frac{c}{\omega}\left[-\beta\ sgn\left(z-z'\right){\bf Q}_{0}+Q{\bf z}\right]$ ( where ${\bf Q}_0$ is the unit vector in the ${\bf Q}$ direction) represent directions of ${\bf s}$(TE) and ${\bf p}$(TM) polarized fields, respectively.        

\end{widetext}

We see that the $z$ integration in (\ref{DysFur}) is restricted exactly from $-L/2$ to $L/2$ which implies that the current fluctuation created in the region $-L/2<z_1<L/2$ can interact via photon propagator ${\bf D}^0({\bf Q},\omega,z_1,z_2)$ (even though the induced electromagnetic filed spreads over the all space) only with the current fluctuation in the region $ L/2<z_2<L/2$. This restriction guarantees that ${\bf\Pi}$ contains information only about the electromagnetic modes characteristic for the electronic system limited to the region $-L/2<z<L/2$ (e.g., Q2D systems). 

The Dyson equation (\ref{DysFur}) can be additionally Fourier transformed in the $z$ direction, so that it becomes a full matrix equation

\begin{eqnarray}
{\bf\Pi}_{{\bf G},{\bf G}'}({\bf Q},\omega)={\bf\Pi}^0_{{\bf G},{\bf G}'}({\bf Q},\omega)+\hspace{3cm}
\nonumber\\
\label{DysFurtot}\\
\sum_{{\bf G}_1,{\bf G}_2}{\bf\Pi}^0_{{\bf G},{\bf G}_1}({\bf Q},\omega)
{\bf D}^0_{{\bf G}_1,{\bf G}_2}({\bf Q},\omega){\bf\Pi}_{{\bf G}_2,{\bf G}'}({\bf Q},\omega),
\nonumber
\end{eqnarray}
where ${\bf G}=({\bf G}_{\parallel},G_z)$ are 3D reciprocal lattice vectors. Here

\begin{equation}
{\bf\Pi}^0_{{\bf G},{\bf G}'}({\bf Q},\omega)={\bf\Pi}^{\mathrm{para}}_{{\bf G},{\bf G}'}({\bf Q},\omega)+{\bf\Pi}^{\mathrm{dia}}_{{\bf G},{\bf G}'}({\bf Q},\omega)
\label{jtu}
\end{equation}
represents the Fourier transform of the current-current response tensor (\ref{current-current}), and where the full Fourier transform of photon propagator can be obtained using (\ref{photpar}), (\ref{photQ}) and 

\begin{eqnarray} 
{\bf D}^0_{{\bf G},{\bf G}'}({\bf Q},\omega)=\hspace{5cm}
\nonumber\\
\frac{1}{L}\int^{L/2}_{-L/2}dzdz'e^{-iG_zz}{\bf D}^0_{{\bf G}_{\parallel},{\bf G}'_\parallel}({\bf Q},\omega,z,z')e^{iG_z'z'}. 
\label{FurttotD}
\end{eqnarray}
Using (\ref{parcurr}), (\ref{oppo}) and (\ref{Self2}) the Fourier transform of the paramagnetic contribution to the current-current response tensor becomes: 

\begin{eqnarray}
\Pi^{\mathrm{para}}_{\mu\nu,{\bf G}{\bf G}'}({\bf Q},\omega)&=&%\hspace{5cm}
%\nonumber\\
%\nonumber\\
-\frac{2}{\Omega \mathrm{c}}\sum_{{\bf K},n,m}\ \frac{f_{n{\bf K}}-f_{m{\bf K}+{\bf Q}}}
{\hbar\omega+i\eta+E_{n{\bf K}}-E_{m{\bf K}+{\bf Q}}}
\nonumber\\
&&\times\ j^{\mu}_{n{\bf K},m{\bf K}+{\bf Q}}({\bf G})\ [j^{\nu}_{n{\bf K},m{\bf K}+{\bf Q}}({\bf G}')]^*,
\label{Pipara}
\end{eqnarray}  
where the current vertices are 

\begin{equation}
j^{\mu}_{n{\bf K},m{\bf K}+{\bf Q}}({\bf G})=
\int_\Omega\ d{\bf r}e^{-i({\bf Q}+{\bf G}_\parallel)\cdot{\brho}-iG_zz}\ 
j^{\mu}_{n{\bf K},m{\bf K}+{\bf Q}}({\bf r}),
\label{curveti}
\end{equation}
and where 

\begin{eqnarray}
j^{\mu}_{n{\bf K},m{\bf K}+{\bf Q}}({\bf r})&=&
\frac{\hbar e}{2im}
\left\{\phi_{n{\bf K}}^*({\bf r})\partial_\mu\phi_{m{\bf K}+{\bf Q}}({\bf r})\right.
\nonumber\\
&&-\left.[\partial_\mu\phi_{n{\bf K}}^*({\bf r})]\phi_{m{\bf K}+{\bf Q}}({\bf r})\right\}.
\label{curvetij}
\end{eqnarray}
Here the 3D position vector is ${\bf r}=(\brho,z)$. Using (\ref{densop}) and (\ref{oppo}) the Fourier transform of the diamagnetic contribution to the current-current response tensor becomes

\begin{equation}
\Pi^{\mathrm{dia}}_{\mu\nu,{\bf G}{\bf G}'}({\bf Q})=
-\delta_{\mu\nu}\frac{2e^2}{mc\Omega}\sum_{{\bf K},n}\ f_n({\bf K})\ \rho_{n{\bf K},n{\bf K}}({\bf G}-{\bf G}')
\label{Furdia}
\end{equation}  
where the charge vertices are  

\begin{equation}
\rho_{n{\bf K},m{\bf K}+{\bf Q}}({\bf G})=
\int_\Omega\ d{\bf r}e^{-i({\bf Q}+{\bf G}_\parallel)\cdot{\brho}-iG_zz}\ \phi_{n{\bf K}}^*({\bf r})\phi_{m{\bf K}+{\bf Q}}({\bf r}).
\label{charverti}
\end{equation} 
Here $\Omega=S\times L$ is the normalization volume, $S$ is the normalization surface and $f_{n{\bf K}}=(e^{(E_{n{\bf K}} E_F)/kT}+1)^{-1}$ is the Fermi-Dirac distribution at temperature $T$. Integrations in (\ref{curveti}) and (\ref{charverti}) are performed over the normalization volume $\Omega$. Plane wave expansion of the wave function has the form 

\[
\Phi_{n{\bf K}}(\brho,z)=\frac{1}{\sqrt{\Omega}}e^{i{\bf K}\cdot\brho}\ \sum_{\bf G}C_{n{\bf K}}({\bf G})e^{i{\bf G}\cdot{\bf r}},
\]
where the coefficients $C_{n{\bf K}}$ are obtained by solving the KS equations self-consistently.

\subsection{Optical absorption spectrum and conductivity tensor}
\label{Abscond}
If the layered system is in interaction with external electromagnetic field described by the vector potential ${\bf A}^{\mathrm{ext}}({\bf r},t)$, the power at which the external electromagnetic energy is absorbed in the system can be obtained from the classical expression 

\begin{equation}
P(t)=\int\ d{\bf r}_1{\bf E}^{\mathrm{ext}}({\bf r}_1,t)\cdot{\bf j}^{\mathrm{ind}}({\bf r}_1,t),
\label{powloss}
\end{equation}
where in the $\Phi=0$ gauge the external electrical field can be calculated from 

\begin{equation} 
{\bf E}^{\mathrm{ext}}=-\frac{1}{\mathrm{c}}\frac{\partial{\bf A}^{\mathrm{ext}}}{\partial t}.
\label{EvsA}
\end{equation}
After inserting the induced current (\ref{zaj}) into (\ref{powloss}) the absorption power becomes \cite{Benzene}

\begin{eqnarray}
P(t)=\hspace{5cm}
\nonumber\\
\label{powerloss}\\
\int^{\infty}_{-\infty}dt_1\int d{\bf r}_1d{\bf r}_2\ {\bf E}^{\mathrm{ext}}({\bf r}_1,t)\mathbf{\Pi}({\bf r}_1,{\bf r}_2,t-t_1){\bf A}^{\mathrm{ext}}({\bf r}_2,t_1).
\nonumber
\end{eqnarray}
Suppose now that the crystal layers plane lie parallel to the $xy$ plane, as shown in Fig. \ref{Fig2}(a), and the incident electromagnetic field is a plane wave of unit amplitude 
and polarization $\mathbf{e}$:
\begin{equation}
{\bf E}^{\mathrm{ext}}({\bf r},t)={\bf e}\cos({\bf q}\cdot{\bf r}-\omega t),  
\label{incidentemp}
\end{equation}
where the incident wave vector is ${\bf q}=({\bf Q},\beta)$ and ${\bf Q}=(Q_{0x},Q_{0y})$ is the wave vector parallel to the $xy$ plane. The dispersion relation for electromagnetic waves in vacuum $\omega=|{\bf q}|\mathrm{c}$ leads to the relation $\beta=\mathrm{\sqrt{\frac{\omega^2}{c^2}-Q^2}}$ where  $\mathrm{Q}=|{\bf Q}|$. This implies that for $\omega>\mathrm{Qc}$ the perturbing field has radiative character with respect to the z-axis, as sketched in Fig. \ref{Fig2}(b), and for $\omega<\mathrm{Qc}$ it has evanescent character as sketched in Fig. \ref{Fig2}(c). The radiative field can excite optically active modes, such as bright excitons, while the evanescent field is suitable for excitation of collective modes such as polaritons. The unit vector ${\bf e}$ represents the polarization of incident electromagnetic field. After combining Eqs. (\ref{EvsA}), (\ref{powerloss}) and (\ref{incidentemp}), we do the Fourier transform of the current-current response tensor ${\bf \Pi}(\omega)=\int^{\infty}_{-\infty}dt e^{i\omega(t-t')}{\bf \Pi}(t-t')$ and the expression for the absorption power becomes

\begin{eqnarray}
P(\omega)=\hspace{5cm}
\nonumber\\
\label{apspower}\\
\frac{c}{2\omega}\mathrm{Im}\left\{\sum_{\mu,\nu}e_\mu e_\nu\int\ 
d{\bf r}_1{\bf r}_2e^{-i{\bf q}\cdot{\bf r}_1}\Pi_{\mu\nu}({\bf r}_1,{\bf r}_2,\omega)e^{i{\bf q}\cdot{\bf r}_2}\right\}. 
\nonumber
\end{eqnarray}
Fourier transforming equations (\ref{zaj}) and (\ref{EvsA}) in $\omega$ space and combining them we obtain 

\begin{equation}
j_\mu^{\mathrm{ind}}({\bf r},\omega)=-i\frac{c}{\omega}\sum_\nu\int\ d{\bf r}_1\Pi_{\mu\nu}({\bf r},{\bf r}_1,\omega)E_\nu^{\mathrm{ext}}({\bf r}_1,\omega).
\label{zajjo}
\end{equation}    
Using the fact that the induced current can also be calculated in terms of electrical conductivity tensor as 

\begin{equation}
{\bf j}_\mu^{\mathrm{ind}}=\sum_{\nu}\sigma_{\mu\nu}\otimes E_{\nu}^{\mathrm{ext}}, 
\label{dosta}
\end{equation}
we obtain the useful connection between the conductivity tensor and current-current response tensor
     
\begin{equation}
\sigma_{\mu\nu}({\bf r},{\bf r}',\omega)=-i\frac{c}{\omega}\Pi_{\mu\nu}({\bf r},{\bf r}',\omega).
\label{condut}
\end{equation}
Now we want to exploit our results  for the tensor $\Pi_{\mu\nu,{\bf G}{\bf G}'}({\bf Q},\omega)$. Fourier transforming it to the real space:

\begin{eqnarray}
\Pi_{\mu\nu}({\bf r},{\bf r}',\omega)=\hspace{5cm}
\nonumber\\
\label{Preal}\\
\frac{1}{L}\sum_{\textbf{G},\textbf{G}'}\int\frac{d{\bf Q}}{(2\pi)^2}
e^{i(\textbf{Q}+\textbf{G})\cdot{\bf r}}e^{-i(\textbf{Q}+\textbf{G}')\cdot{\bf r}'}
\Pi_{\mu\nu,{\bf G}{\bf G}'}({\bf Q},\omega)
\nonumber
\end{eqnarray}
and inserting in (\ref{apspower}) the absorption power per unit area becomes

\begin{equation}
P({\bf Q},\omega)=\frac{c}{2\omega L}S({\bf Q},\omega)
\label{absPFT}
\end{equation}
where the spectral function is

\begin{eqnarray}
S({\bf Q},\omega)=\hspace{5cm}
\nonumber\\
\mathrm{Im}\left\{\sum_{\mu,\nu}e_\mu e_\nu\sum_{G_z,G_z'}
F(G_z)F(G_z')\Pi_{\mu\nu,G_zG_z'}({\bf Q},\omega)\right\}
\label{spectrfun}
\end{eqnarray}
and the form factors are $F(G_z)=\frac{2}{G_z-\beta}\sin\frac{(G_z-\beta)L}{2}$. It is more convenient to deal with absorption coefficient $A({\bf Q},\omega)=P({\bf Q},\omega)/|{\cal P}|$ where the incident flux of electromagnetic energy (Poynting vector) is given by ${\cal P}=\frac{c}{4\pi}{\bf E}\times{\bf B}$. For unit amplitude incident electrical field (\ref{incidentemp}) the incident flux is $|{\cal P}|=\frac{c}{8\pi}$ and the absorption coefficient is
\begin{equation}
A({\bf Q},\omega)=\frac{4\pi}{\omega L}S({\bf Q},\omega).
\label{abscoef}
\end{equation}
The Fourier transform of the conductivity tensor (\ref{condut}) can be obtained directly from the Fourier transform of current-current response tensor as

\begin{equation}
\sigma_{\mu\nu,{\bf G}{\bf G}'}({\bf Q},\omega)=-i\frac{c}{\omega}\Pi_{\mu\nu,{\bf G}{\bf G}'}({\bf Q},\omega).
\label{FTcondut}
\end{equation}
The current which is induced by a homogeneous electric field ${\bf E}={\bf e}\cos\omega t$ (which corresponds to electromagnetic field (\ref{incidentemp})) which is incident normally (${\bf Q}=0$) and for $\beta\ll 1/L$ can be, using (\ref{dosta}) and (\ref{FTcondut}), written as    

\begin{equation}
j_\mu^{\mathrm{ind}}(z,\omega)=\frac{c}{\omega}\sum_{G_z}e^{iG_zz}\sum_\nu \mathrm{Im}\Pi_{\mu\nu,G_z0}({\bf Q}=0,\omega)e_\nu. 
\label{gstr}
\end{equation}
Here, as we are not interested in the current variation within the unit cell in the parallel $xy$ direction, we retained only  ${\bf G}_\parallel={\bf G}'_\parallel=0$ components. If the field is directed along the $x$ or $y$ axis the current flow per unit thickness of the sample can be obtained by $z$ integration in (\ref{gstr}), when it becomes:

\begin{equation}
\mathrm{Re}\sigma_{\mu\mu}(\omega)=\frac{cL}{\omega}\ \mathrm{Im}\Pi_{\mu\mu,00}({\bf Q}=0,\omega);\ \ \mu=x,y, 
\label{sigmaexp}
\end{equation}
which corresponds to the experimentally measurable conductivity.

\subsection{Alternative expression for the current-current response tensor}
\label{HAlter}
For numerical reasons a straightforward calculation of the current-current response tensor (\ref{jtu}) from Eqs. (\ref{Pipara}--\ref{charverti}) can lead to non-physical results, so here we shall derive an alternative expression which avoids this problem. Namely, the expression for $\Pi^{\mathrm{para}}$ (\ref{Pipara}) includes summation over all unoccupied bands, which is not the case for $\Pi^{\mathrm{dia}}$ (\ref{Furdia}), so if the calculation is not performed with the same high precision the result for $\Pi$ (\ref{jtu}) might be erroneous.  

Suppose for the moment that the system interacts only with the external electromagnetic field ${\bf A}^{\mathrm{ext}}({\bf r},t)$ and that the interaction $V^{\mathrm{in}}$ is neglected. Then the induced current and charge distributions are given by (\ref{zaj}) and (\ref{roire}), respectively. After inserting (\ref{zaj}) and (\ref{roire}) into the continuity equation (\ref{nabcon}) and performing the Fourier transform in $({\bf q},\omega)$ space it becomes \cite{Schrieffer}

\begin{equation}
\omega\Pi_{0\nu,{\bf G}{\bf G}'}^0({\bf Q},\omega)=\sum_\mu\ q_\mu\Pi^0_{\mu\nu,{\bf G}{\bf G}'}({\bf Q},\omega),
\label{fteqc}
\end{equation}
where $(q_x,q_y,q_z)=({\bf Q}+{\bf G}_\parallel,G_z)$ and $\nu=x,y,z$. 
The Fourier transform of the current-current response tensor $\Pi_{\mu\nu}$ is given by (\ref{jtu}), (\ref{Pipara}) and (\ref{Furdia}) and the Fourier transform of the charge-current response tensor $\Pi_{0\nu}$, defined by (\ref{charcurt}), is given by

\begin{eqnarray}
\Pi_{0\nu,{\bf G}{\bf G}'}^0({\bf Q},\omega)&=&
-\frac{2}{\Omega \mathrm{c}}\sum_{{\bf K},n,m}\ \frac{f_{n{\bf K}}-f_{m{\bf K}+{\bf Q}}}
{\hbar\omega+i\eta+E_{n{\bf K}}-E_{m{\bf K}+{\bf Q}}}
\nonumber\\
&&\times
\rho_{n{\bf K},m{\bf K}+{\bf Q}}({\bf G})\ [j^{\nu}_{n{\bf K},m{\bf K}+{\bf Q}}({\bf G}')]^*,
\label{Pirocur}
\end{eqnarray}
where the current and charge vertices are defined by (\ref{curveti}) and (\ref{charverti}), respectively. In the static limit ($\omega=0$), when the current flow becomes stationary (also called the direct current or DC limit) the continuity equation (\ref{fteqc}) becomes 

\begin{equation}
\sum_\mu\ q_\mu\Pi^0_{\mu\nu,{\bf G}{\bf G}'}({\bf Q},0)=0.
\label{stateqc}
\end{equation}
It is important to note that $\Pi_{\mu\nu}^0$ should be calculated so that condition (\ref{stateqc}) is satisfied very accurately, otherwise it can lead to some incorrect physical conclusions. For example, using the definition of the conductivity tensor (\ref{condut}), the DC conductivity becomes  

\begin{equation}
\sigma_{\mu\nu,{\bf G}{\bf G}'}({\bf Q},\omega\rightarrow 0)=-ic\lim_{\omega\rightarrow 0}\frac{\Pi_{\mu\nu,{\bf G}{\bf G}'}({\bf Q},\omega)}{\omega}.
\label{DCsig}
\end{equation}
We see that if the condition (\ref{stateqc}) is not satisfied (\ref{DCsig}) could lead to the wrong conclusion about, e.g., the existence of superconducting state ($\sigma\rightarrow\infty$). Also, in the normal metal state it could affect the Drude plasmon frequency. As already mentioned, the problem arises from the numerical calculation of the paramagnetic term (\ref{Pipara}) which includes summation over all unoccupied bands and will never be capable to cancel exactly the diamagnetic contribution (\ref{Furdia}) which includes only the summation over occupied states and can be calculated very accurately. 

This problem can be solved so that we calculate the paramagnetic term at some appropriate level of accuracy and then require the diamagnetic term to satisfy the continuity equation (\ref{stateqc}), i.e.    

\begin{equation}
\Pi^{\mathrm{dia}}_{\mu\nu,{\bf G}{\bf G}'}({\bf Q})=-\Pi^{\mathrm{para}}_{\mu\nu,{\bf G}{\bf G}'}({\bf Q},0).
\label{dianew}
\end{equation}
After using (\ref{jtu}), (\ref{Pipara}) and (\ref{dianew}) the redefined current-current response tensor becomes 

\begin{eqnarray}
\Pi^0_{\mu\nu,{\bf G}{\bf G}'}({\bf Q},\omega)&=&
\frac{2}{\Omega \mathrm{c}}\sum_{{\bf K},n,m}\ \frac{\hbar\omega}{E_{n{\bf K}}-E_{m{\bf K}+{\bf Q}}}
\nonumber\\
&&\times
\frac{f_{n{\bf K}}-f_{m{\bf K}+{\bf Q}}}
{\hbar\omega+i\eta+E_{n{\bf K}}-E_{m{\bf K}+{\bf Q}}}
\nonumber\\
&&\times
j^{\mu}_{n{\bf K},m{\bf K}+{\bf Q}}({\bf G})\ [j^{\nu}_{n{\bf K},m{\bf K}+{\bf Q}}({\bf G}')]^*.
\label{Pitotnew}
\end{eqnarray}  

Now we shall demonstrate that the current-current response tensor given by (\ref{Pitotnew}) satisfies the continuity equation in the whole frequency range, i.e. it satisfies the equation (\ref{fteqc}). We start from the operator form of the continuity equation, which for the system of independent electrons and without interaction with the external field can be written as 

\begin{equation}
[\rho({\bf r},t),H_0^{e}]=\frac{i}{\hbar}\nabla{\bf j}^{\mathrm{para}}({\bf r},t)
\label{opeqcon}
\end{equation}
where $\hat{O}(t)=\left\{\rho(t),{\bf j}(t)\right\}$ are Heisenberg operators, defined as $\hat{O}(t)=e^{iH_0^{e}t}\hat{A}e^{iH_0^{e}t}$. The Schr\"odinger operators $\hat{O}=\left\{\rho,{\bf j}\right\}$ are defined as (\ref{parcurr}) and (\ref{densop}) and the Hamiltonian $H_0^{e}$ is given by (\ref{ineham}). After Fourier transformation of equation (\ref{opeqcon}), i.e. $\int d{\bf r}e^{-i{\bf q}{\bf r}}\left\{[\rho({\bf r},t),H_0^{e}]=\frac{i}{\hbar}\nabla{\bf j}({\bf r},t)\right\}$, and some commutation relations manipulation it becomes: 

\begin{eqnarray}
&&\sum_{{\bf K},n,m}\rho_{n{\bf K},m{\bf K}+{\bf Q}}({\bf G})\left[E_{n{\bf K}}-E_{m{\bf K}+{\bf Q}}\right]c^{\dag}_{n{\bf K}}c_{m{\bf K}+{\bf Q}}=
\nonumber\\
&&-\sum_{{\bf K},n,m}\sum_{\mu}\hbar q_\mu\ j^{\mu}_{n{\bf K},m{\bf K}+{\bf Q}}({\bf G})c^{\dag}_{n{\bf K}}c_{m{\bf K}+{\bf Q}}
\label{ftoeqcon}
\end{eqnarray}  
By equating left and right sides in (\ref{ftoeqcon}) we obtain a useful connection between the charge and current vertices
\begin{equation}
\rho_{n{\bf K},m{\bf K}+{\bf Q}}({\bf G})=
-\sum_{\mu}\hbar q_\mu\ \frac{j^{\mu}_{n{\bf K},m{\bf K}+{\bf Q}}({\bf G})}{E_{n{\bf K}}-E_{m{\bf K}+{\bf Q}}}.
\label{rhojcon}
\end{equation}   
After inserting charge vertices (\ref{rhojcon}) into charge-current response function (\ref{Pirocur}) and then into the left hand 
side of the continuity equation (\ref{fteqc}) it becomes exactly equal to its right hand side in which the current-current response tensor (\ref{Pitotnew}) is inserted. Therefore, current-current response tensor (\ref{Pitotnew}) satisfies the continuity equation for any $\omega$.

Important aspect of the new current-current response tensor (\ref{Pitotnew}) is appearance of the $\hbar\omega/\left(E_{n{\bf K}}-E_{m{\bf K}+{\bf Q}}\right)$ prefactor which ensures that $\Pi^0_{\mu\nu}({\bf Q},\omega)\rightarrow0$ when $\omega\rightarrow0$ and also naturally compensates the $\omega^{-1}$ divergence in the Kubo conductivity formula.

\subsection{The long wavelength limit, $\mathbf{Q}\approx0$}
\label{dugi}
We shall now analyze the long wavelength limit, $\mathbf{Q}\approx0$, of the redefined current-current response tensor (\ref{Pitotnew}). First it is convenient to decompose the current-current response tensor into its intraband ($n=m$) and interband ($n\neq m$) contributions,

\begin{equation}
\Pi^0_{\mu\nu,{\bf G}{\bf G}'}({\bf Q},\omega)=\Pi^{0,\mathrm{intra}}_{\mu\nu,{\bf G}{\bf G}'}({\bf Q},\omega)+\Pi^{0,\mathrm{inter}}_{\mu\nu,{\bf G}{\bf G}'}({\bf Q},\omega)
\label{eq1}
\end{equation}
For $\mathbf{Q}\approx0$ and $n=m$ we have that $E_{n{\bf K}}-E_{m{\bf K}+{\bf Q}}\approx0$, and we can write the intraband contribution as \cite{Kupcic,Kupcic1},

\begin{eqnarray}
\Pi^{0,\mathrm{intra}}_{\mu\nu,{\bf G}{\bf G}'}(\omega)&=&
\frac{2}{\Omega \mathrm{c}}\frac{\hbar\omega}{\hbar\omega+i\eta_{\mathrm{intra}}}\sum_{{\bf K},n}\frac{\partial f(E_{n{\bf K}})}{\partial E_{n{\bf K}}}
\nonumber\\
&&\times
j^{\mu}_{n{\bf K},n{\bf K}}({\bf G})\ [j^{\nu}_{n{\bf K},n{\bf K}}({\bf G}')]^*,
\label{eq2}
\end{eqnarray}  
where we changed $f_{n{\bf K}}\rightarrow f(E_{n{\bf K}})$, and for simplicity we write $\Pi^{0}_{\mu\nu,{\bf G}{\bf G}'}({\bf Q}\approx0,\omega)\equiv\Pi^{0}_{\mu\nu,{\bf G}{\bf G}'}(\omega)$. This intraband term leads to the Drude conductivity formula, as will be shown below. For $n\neq m$ we have,

\begin{eqnarray}
\Pi^{0,\mathrm{inter}}_{\mu\nu,{\bf G}{\bf G}'}(\omega)&=&
\frac{2}{\Omega \mathrm{c}}\sum_{{\bf K},n\neq m}\ \frac{\hbar\omega}{E_{n{\bf K}}-E_{m{\bf K}}}
\nonumber\\
&&\times
\frac{f_{n{\bf K}}-f_{m{\bf K}}}
{\hbar\omega+i\eta_{\mathrm{inter}}+E_{n{\bf K}}-E_{m{\bf K}}}
\nonumber\\
&&\times
j^{\mu}_{n{\bf K},m{\bf K}}({\bf G})\ [j^{\nu}_{n{\bf K},m{\bf K}}({\bf G}')]^*.
\label{eq3}
\end{eqnarray}  
Using some simple manipulation we can bring the expression (\ref{eq3}) into the following form \cite{Kupcic2},

\begin{eqnarray}
\Pi^{0,\mathrm{inter}}_{\mu\nu,{\bf G}{\bf G}'}(\omega)&=&
\frac{2}{\Omega \mathrm{c}}\sum_{{\bf K},n\neq m}\ \frac{(\hbar\omega)^2}{E_{n{\bf K}}-E_{m{\bf K}}}
\nonumber\\
&&\times
\frac{f_{n{\bf K}}-f_{m{\bf K}}}
{\hbar\omega(\hbar\omega+2i\eta_{\mathrm{inter}})-\left(E_{n{\bf K}}-E_{m{\bf K}}\right)^2}
\nonumber\\
&&\times
j^{\mu}_{n{\bf K},m{\bf K}}({\bf G})\ [j^{\nu}_{n{\bf K},m{\bf K}}({\bf G}')]^*.
\label{eq4}
\end{eqnarray}  
One can easily see that the above expression for the interband term has a different behaviour in the static limit ($\omega=0$) than expression (\ref{eq3}). The expressions (\ref{eq2}), (\ref{eq3}) and (\ref{eq4}) will be used to calculate the adsorption coefficient (\ref{abscoef}) and conductivity (\ref{sigmaexp}) for optical wavevectors $\mathbf{Q}\sim\mathbf{Q}_{\mathrm{light}}$.

In addition, we shall verify that (\ref{Pitotnew}) leads to the correct dielectric function and conductivity in the three dimensional electron gas in the long wavelength ($\mathbf{q}\approx0$) limit.  For a polarisable system of arbitrary symmetry (in the linear response approximation) the electric field ${\bf E}$ and electric displacement ${\bf{\cal D}}$ can be related as    

\begin{equation}
{\bf{\cal D}}=\beps\otimes{\bf E} 
\label{defeps}
\end{equation}
where $\beps$ is the non-local dielectric tensor and $\otimes$ is matrix multiplication and convolution in real space. After combining the definition (\ref{defeps}) with the Maxwell and Dyson equations for electrical field ${\bf E}$ in the presence of a polarisable system we can obtain a general relationship between the dielectric tensor and the current-current response tensor \cite{Pol,Casimir}

\begin{equation} 
\varepsilon_{\mu\nu}({\bf r},{\bf r}',\omega)=\delta({\bf r}-{\bf r}')\delta_{\mu\nu}+
\frac{4\pi c}{\omega^2}
\Pi^0_{\mu\nu}({\bf r},{\bf r}',\omega).
\label{teneps0}
\end{equation}
If we consider a 3D homogeneous electron gas the Fourier transform of (\ref{teneps0}) can be written as

\begin{equation} 
\varepsilon_{\mu\nu}({\bf q},\omega)=\delta_{\mu\nu}+
\frac{4\pi c}{\omega^2}
\Pi^0_{\mu\nu}({\bf q},\omega),
\label{teneps}
\end{equation}
where $\Pi^0_{\mu\nu}({\bf q},\omega)$ can be obtained from (\ref{Pitotnew}), ${\bf G}={\bf G}'=0$, ${\bf K}\rightarrow{\bf k}$ becomes a 3D wave vector and ${\bf Q}\rightarrow{\bf q}$ becomes 3D momentum transfer. Also, after using the fact that in a homogeneous electron gas there is only one ($n=m=1$) parabolic band ($E_{\bf k}=\frac{\hbar^2{\bf k}^2}{2m}$) and the wave functions are plane waves ($\Phi_{{\bf k}}({\bf r})=\frac{1}{\sqrt{\Omega}}e^{i{\bf k}{\bf r}}$) the current-current response tensor (\ref{Pitotnew}) becomes  

\begin{eqnarray}
\Pi^0_{\mu\nu}({\bf q},\omega)&=&
\frac{\hbar^3\omega e^2}{2m^2\Omega \mathrm{c}}\sum_{{\bf k}}\frac{(2k_\mu+q_\mu)(2k_\nu+q_\nu)}{E_{\bf k}-E_{{\bf k}+{\bf q}}}
\nonumber\\
&&\times
\frac{f_{{\bf k}}-f_{{\bf k}+{\bf q}}}
{\hbar\omega+i\eta+E_{{\bf k}}-E_{{\bf k}+{\bf q}}},
\end{eqnarray} 
where we also used the definition of the current vertices (\ref{curveti}) and (\ref{curvetij}). One can easily show that such a current-current response tensor in the long wavelength limit becomes 

\begin{equation}
\Pi^0_{\mu\nu}({\bf q}\approx 0,\omega)=-\frac{ne^2}{mc}\ \frac{\omega}{\omega+i\eta}\delta_{\mu\nu}
\label{drudePi}
\end{equation}
where $n=\frac{1}{\Omega}\sum_{\bf k}f({\bf k})$ is the electron density. After inserting (\ref{drudePi}) into (\ref{teneps}) we get the well known Drude dielectric function: 

\begin{equation} 
\varepsilon_{\mu\nu}({\bf q}\approx 0,\omega)=\left[1-\frac{\omega_p^2}{\omega(\omega+i\eta)}\right]\delta_{\mu\nu}
\label{Drudeeps}
\end{equation}
where $\omega_p=\sqrt{4\pi ne^2/m}$ is the bulk plasmon frequency. Similarly, combining (\ref{drudePi}) and the definition (\ref{condut}) leads to the Drude conductivity tensor \cite{Mahknj}

\begin{equation}
\sigma_{\mu\nu}({\bf q}\approx 0,\omega)=i\frac{ne^2}{m}\ \frac{1}{\omega+i\eta}\delta_{\mu\nu}.
\end{equation}
From the definition of the simple Drude DC conductivity $\sigma=\frac{ne^2}{m}\tau$ it is obvious that $\eta$ plays the role of the inverse relaxation time, i.e. $\eta=1/\tau$.

\section{Results and discussion}
\label{Results}
In order to illustrate the advantages of our theoretical approach we shall apply it to calculate optical properties of a free-standing single layer graphene.
\subsection{Computational details}
\label{Comp}
The first part of the calculation consists of determining the KS ground state of the single layer graphene and the corresponding wave functions and energies. For the unit cell constant we use the experimental value of $a=4.651\ \mathrm{a.u.}$ \cite{lattice}, and we separate the graphene layers with the distance $L=5a$. For calculating KS wave functions and energies we use a plane-wave self-consistent field DFT code (PWSCF) within the QUANTUM ESPRESSO (QE) package \cite{QE}. The core-electron interaction was approximated by the norm-conserving pseudopotentials \cite{normcon}, and the exchange correlation (XC) potential by the Perdew-Zunger local density approximation (LDA) \cite{lda1}. To calculate the ground state electronic density we use $30\times30\times1$ Monkhorst-Pack K-point mesh \cite{MPmesh} of the first Brillouin zone (BZ) and for the plane-wave cut-off energy we choose 50 Ry. In order to achieve better resolution in the low energy and the static limit ($\omega\rightarrow 0$) the current response tensor (\ref{Pitotnew}) is evaluated from the wave functions $\phi_{n{\bf K}}({\bf r})$ and energies $E_n({\bf K})$ calculated for the $402\times402\times1$ Monkhorst-Pack K-point mesh, and band summation is performed over 30 bands. In the calculation we use two kinds of damping parameters: $\eta_{\mathrm{intra}}$ for transitions within the same band ($n\rightarrow n$), and $\eta_{\mathrm{inter}}$ for transitions between different bands ($n\rightarrow m$). These two damping energies will be variable parameters. We take $G_\parallel=G'_\parallel=0$ in all the calculations because the crystal local field effects in the crystal layer plane ($xy$) are negligible in the optical limit ($\mathrm{Q}\approx 0$). However, broken symmetry in $z$ direction results in big inhomogeneity of induced currents and fields in that direction. This requires inclusion of the crystal local field effects in $z$ direction which we describe with $21$ $G_z$ Fourier components. After solving the Dyson equation (\ref{DysFurtot}) we obtain the screened current-current response tensor ${\bf\Pi}$ which enters in the absorption coefficient (\ref{abscoef}). The conductivity tensor (\ref{FTcondut}) can be, due to negligible screening for $Q\approx 0$, calculated from the unscreened current-current response tensor (\ref{Pitotnew}).

\begin{figure}[b]
\centering
\includegraphics[width=\columnwidth]{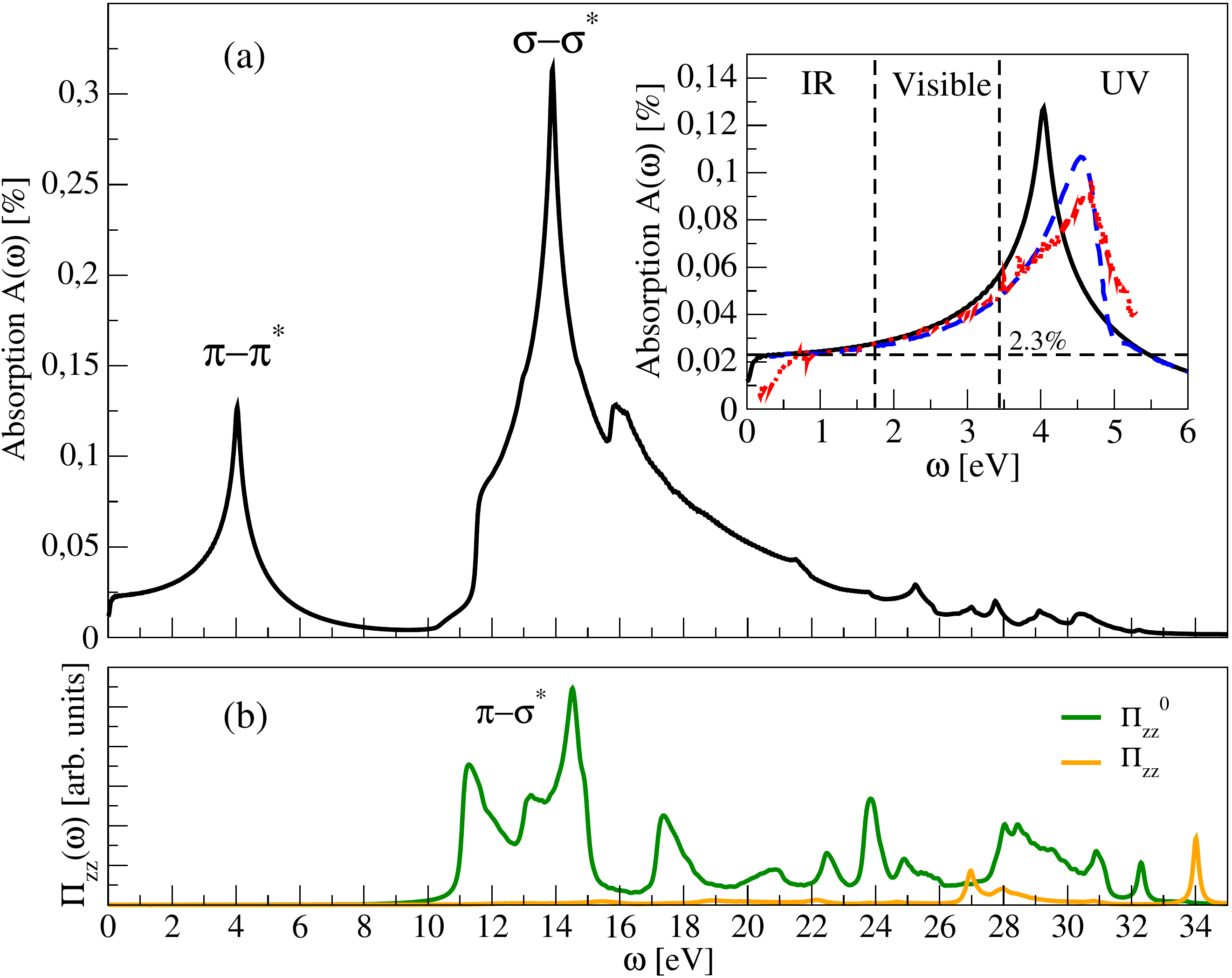}
\caption{(Color online) (a) Absorption spectrum in pristine graphene ($E_F=0$) for normal incidence ($\mathrm{Q}=0$) and $T=0\ \mathrm{K}$. Inset: Details of optical absorption in IR, visible and UV regions; black solid lines: this work, blue dashed lines: theoretical results taken from Ref. \cite{Louie}, and red dotted lines: experimental results taken from Ref. \cite{SSC}. The damping parameters are $\eta_{\mathrm{intra}}=10\ \mathrm{meV}$ and $\eta_{\mathrm{inter}}=50\ \mathrm{meV}$. (b) Non-interacting current-current correlation tensor (\ref{Pitotnew}) (green line) and the interacting one (\ref{DysFurtot}) (yellow line) for $\mu=\nu=z$ and $\mathrm{Q}=0$.}
\label{Fig3}
\end{figure}

\begin{figure}[t]
\centering
\includegraphics[width=8cm,height=6cm]{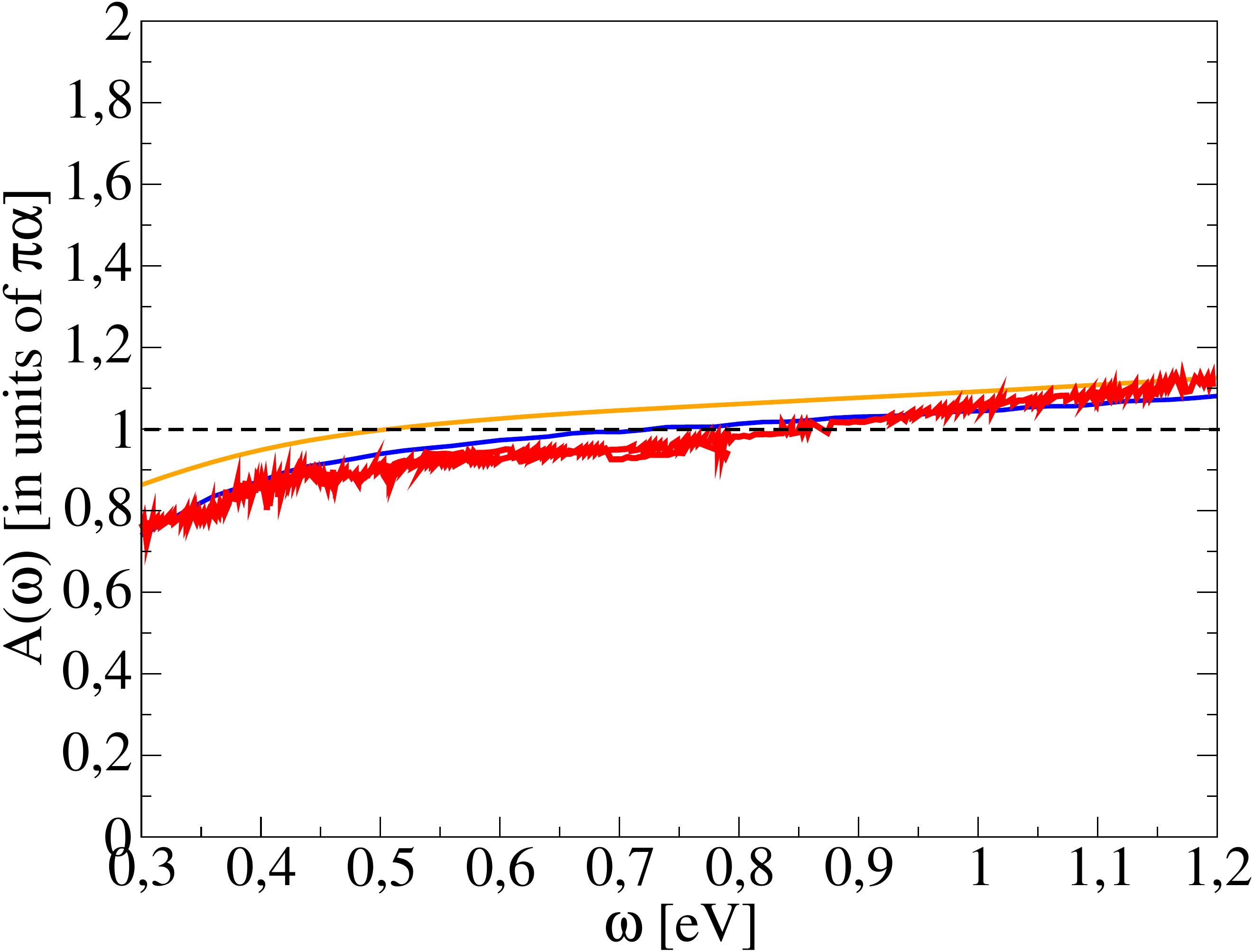}
\caption{(Color online) Calculated absorption spectrum in slightly doped graphene ($E_F=0.1\ \mathrm{eV}$) for normal incidence ($\mathrm{Q}=0$) and $T=300\ \mathrm{K}$ obtained with Eq. (\ref{eq3}) (yellow line) and Eq. (\ref{eq4}) (blue line). Experimental absorption spectrum for graphene sample taken from Ref. \cite{2.3posto2} is represented with red line. Horizontal dashed line denotes the universal absorption constant $\pi\alpha$. The interband damping parameter used in this calculations is $\eta_{\mathrm{inter}}=50\ \mathrm{meV}$.}
\label{Fig33}
\end{figure}

\subsection{Optical absorption} 
\label{lightabs}
In this section we study the absorption of incident electromagnetic field (\ref{incidentemp}) in the free-standing single layer graphene. We fix the parallel component of the incident wavevector ${\bf Q}=(\mathrm{Q}_{0x},\mathrm{Q}_{0y})$ and change the incident frequency $\omega$. Due to the relation $\omega=|{\bf q}|c$, for $\omega<\mathrm{Qc}$ the perpendicular component of incident wavevector $\beta$ is imaginary, and the incident field has evanescent character (in $z$ direction), as sketched in Fig. \ref{Fig2}(b). For $\omega\geq \mathrm{Qc}$ $\beta$ is real and the incident field has radiative character in all three directions [Fig. \ref{Fig2}(c)]. In the latter case the incident wavevector ${\bf q}$ is inclined relative to the graphene surface by an 
angle $\theta$ given by    
\[
\sin\theta=\frac{\beta}{|{\bf q}|}=\sqrt{1-\frac{\mathrm{Q^2c^2}}{\omega^2}},
\]
as sketched in Fig. \ref{Fig2} (a).

Let us discuss some specific cases. For example, for $\omega=\mathrm{Qc}$ (on the light cone) the incident electromagnetic field is a plane wave which propagates parallel to the $xy$ plane ($\theta=0$), ${\bf s}$ polarization is in $xy$ plane, and ${\bf p}$ polarization is wholly in the perpendicular $z$ direction. Also, if $\mathrm{Q}=0$, the electromagnetic field has radiative character in the whole frequency range ($\omega>0$), incidence is normal to the graphene surface ($\theta=\pi/2$) and ${\bf s}$ and ${\bf p}$ polarizations become equal.

\begin{figure*}[!ht]
\includegraphics[width=0.8\columnwidth]{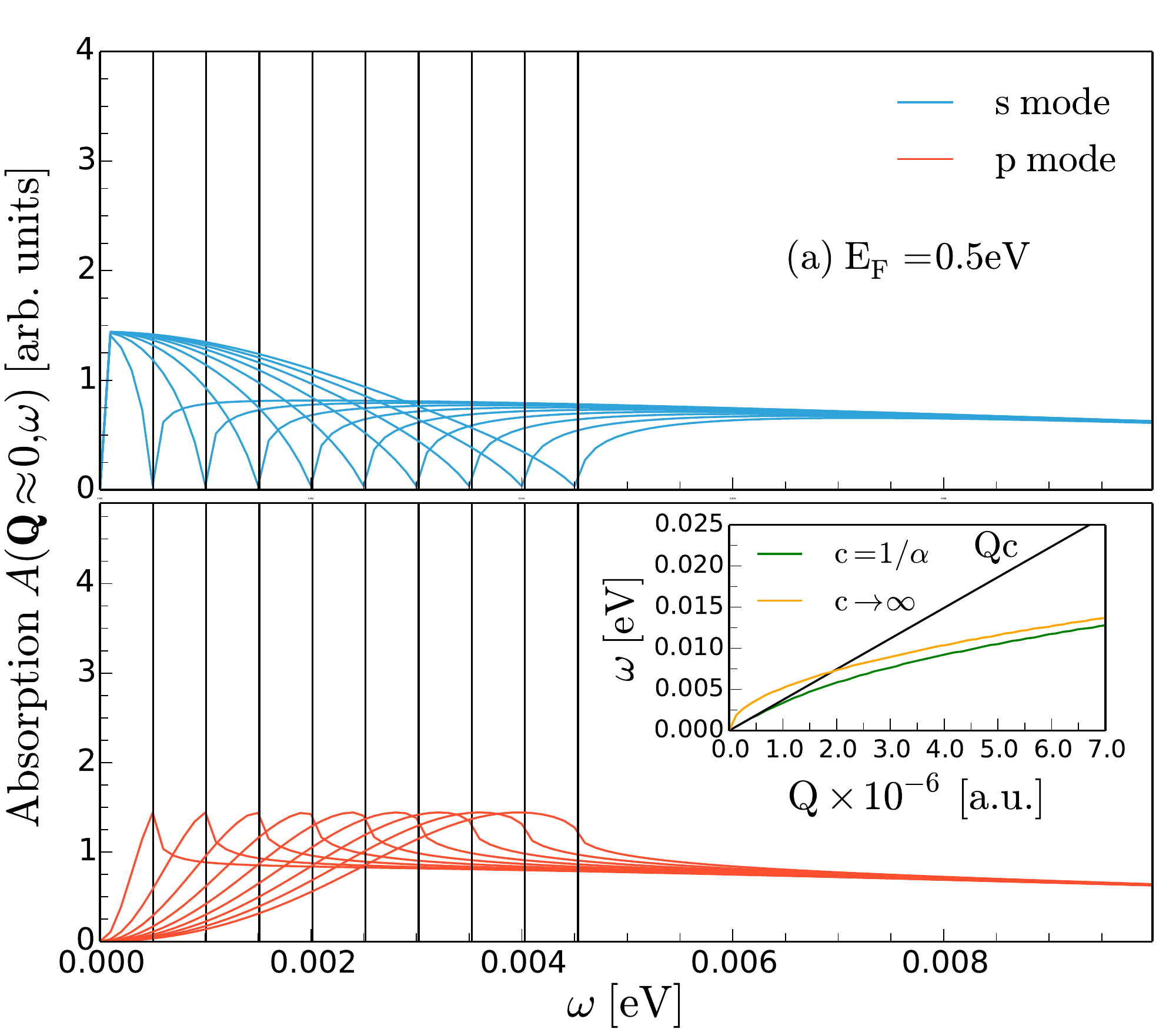}
\includegraphics[width=0.8\columnwidth]{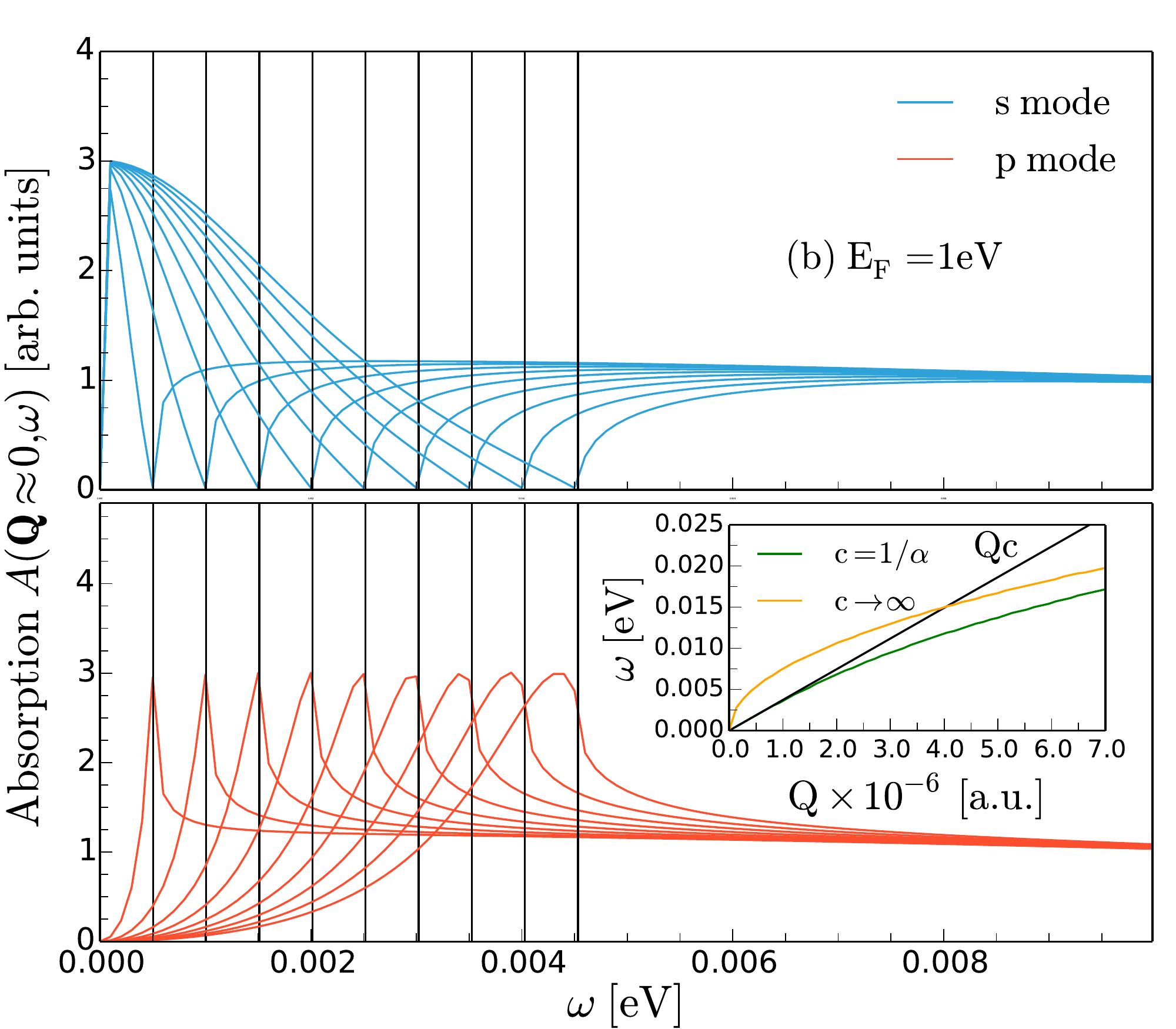}
\caption{(Color online) Absorption spectrum for $\mathbf{s}$ (blue lines) and $\mathbf{p}$ (red lines) mode in doped graphene. Results are presented for (a) $E_F=0.5\ \mathrm{eV}$ and (b) $E_F=1\ \mathrm{eV}$. Incident wavevectors are in $\Gamma\mathrm{M}$ direction with the values ${\bf Q}_n=n\Delta \mathrm{Q}\ {\bf \hat{y}},\ \ n=0-8$, where $\Delta \mathrm{Q}=1.35\times10^{-7}\ \mathrm{a.u.}$. Vertical solid lines represent positions of the energy of light ($\omega_{\mathrm{light}}=\mathrm{Qc}$) for each of the wavevector Q. Insets: Dispersion of the longitudinal 2D plasmon (TM mode) for each of the 
dopings, calculated with $\mathrm{c=1/\alpha}$ (green line) and with $\mathrm{c\rightarrow\infty}$ (yellow line).} 
\label{Fig34}
\end{figure*}

\begin{figure*}[!ht]
\includegraphics[width=0.65\columnwidth]{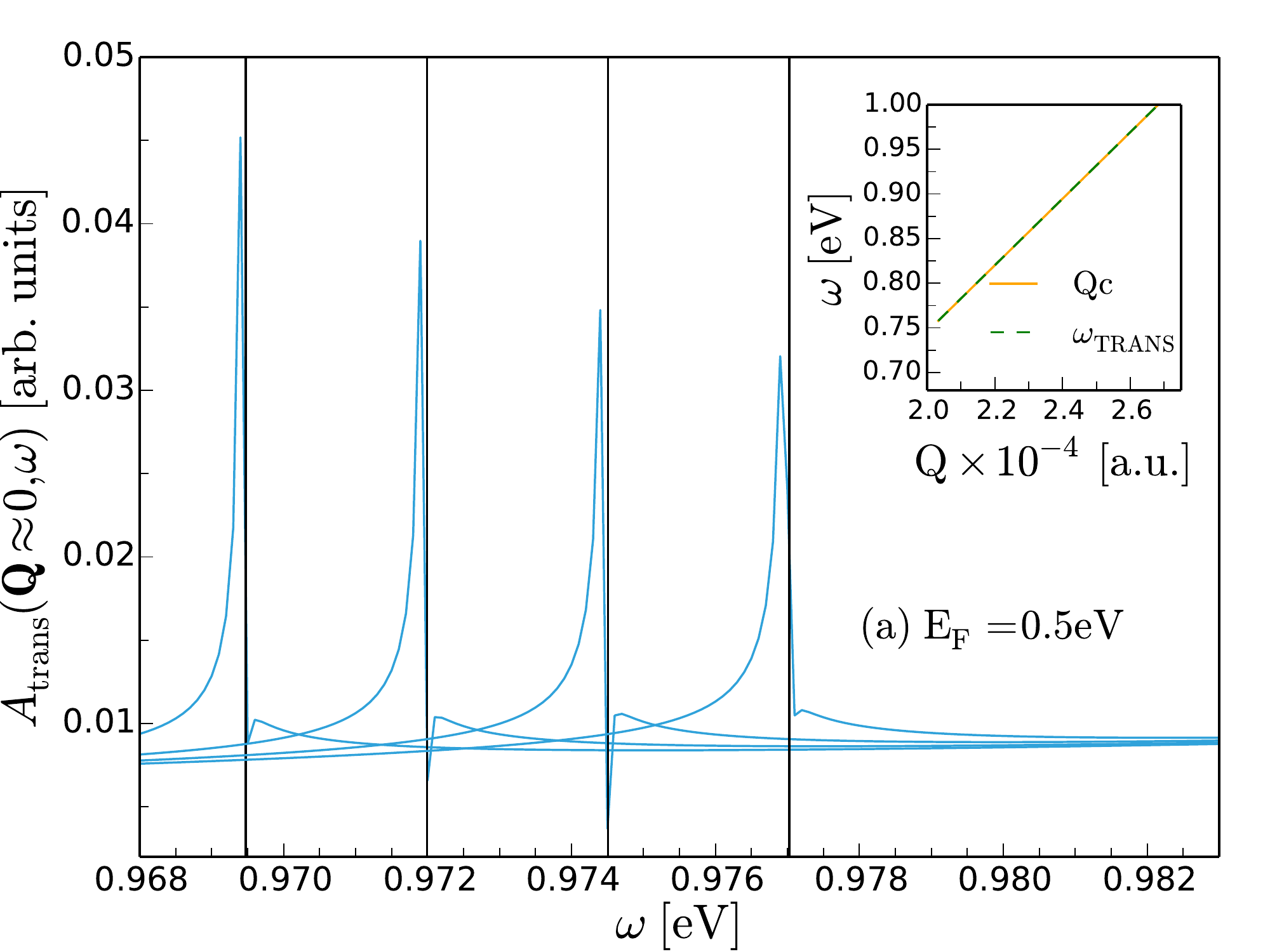}
\includegraphics[width=0.65\columnwidth]{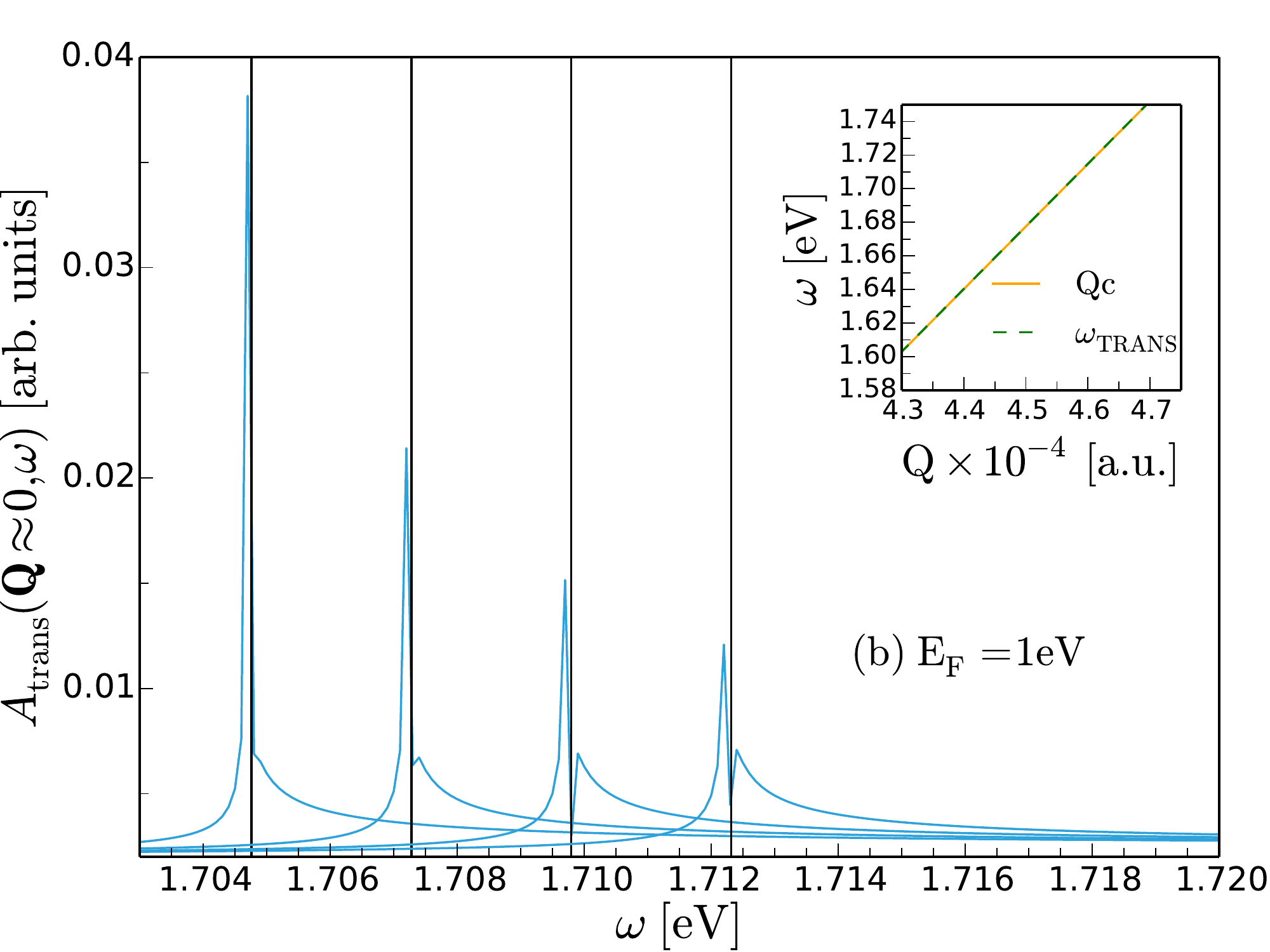}
\includegraphics[width=0.7\columnwidth]{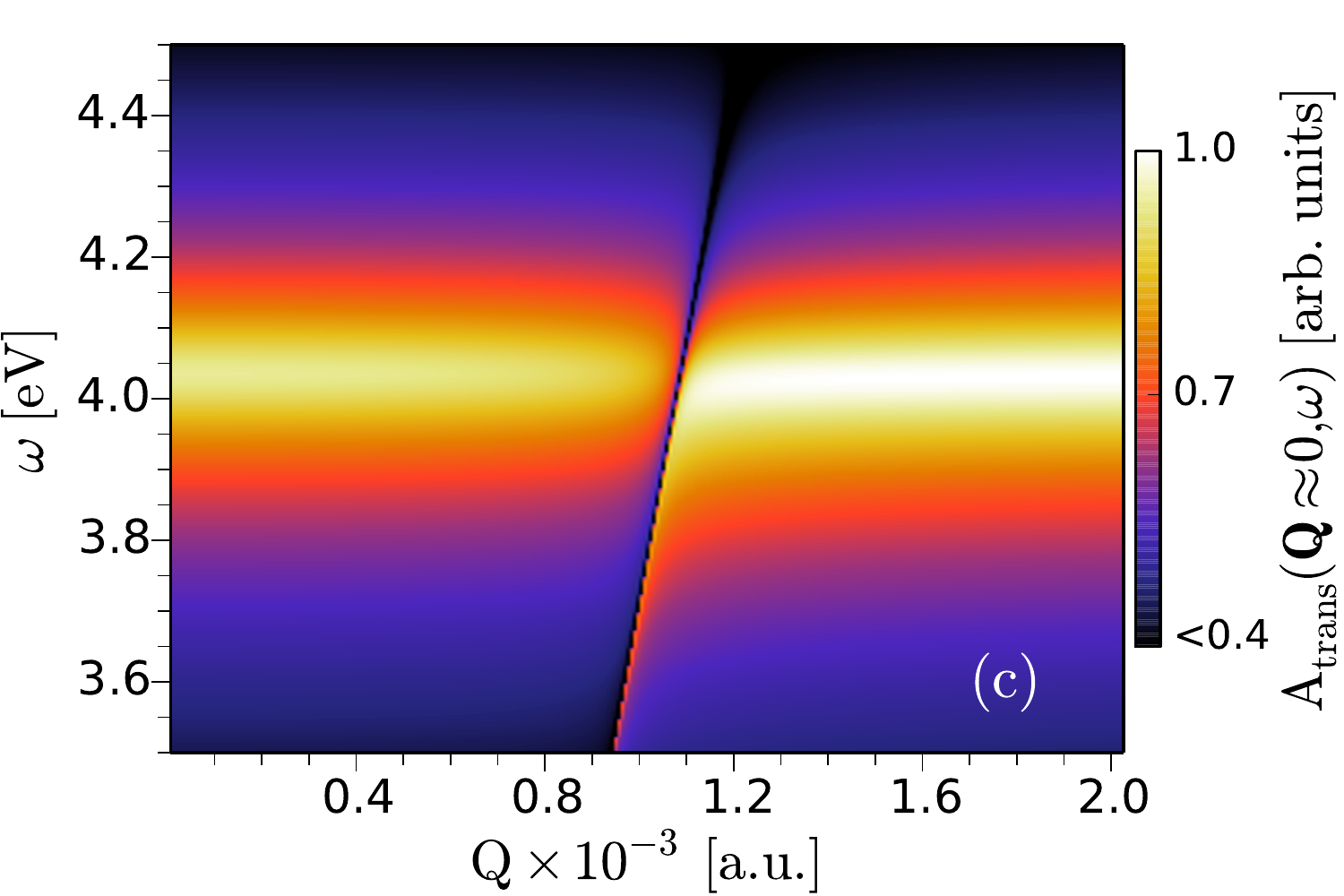}
\caption{(Color online) (a) Absorption spectrum for $\mathbf{s}$ (blue lines) modes in doped graphene ($E_F=0.5\ \mathrm{eV}$) where incident wavevectors are ${\bf Q}_n=\mathrm{Q}_0+n\Delta \mathrm{Q}\ {\bf \hat{y}},\ \ n=0-3$, with $\mathrm{Q_0=2.6\times10^{-4}\ a.u.}$ and $\Delta \mathrm{Q}=6.76\times10^{-7}\ \mathrm{a.u.}$. (b) Same as in (a) but with $E_F=1\ \mathrm{eV}$, $\mathrm{Q_0=4.6\times10^{-4}\ a.u.}$ and $\Delta\mathrm{Q}=6.76\times10^{-7}\ \mathrm{a.u.}$. Vertical solid lines are positions of the energy of light ($\omega_{\mathrm{light}}=\mathrm{Qc}$) for the corresponding wavectors Q. Insets: Dispersion of the TE mode for each of the dopings compared with the dispersion of light. (c) Absorption intensities for $\mathbf{s}$ mode around the energies of $\pi\rightarrow\pi^{\ast}$ transitions for very small Q wavevectors ($\mathrm{Q\sim Q_{light}}$).} 
\label{Fig35}
\end{figure*} 

Retardation effects are most pronounced for very small wavevectors, so we shall divide our discussion of optical absorption into two parts: for small $\mathrm{Q\approx Q_{light}}$, and large $\mathrm{Q\gg Q_{light}}$, wavevectors.

\subsubsection{$\mathrm{Q\approx Q_{light}}$}

\begin{figure*}[!ht]
\includegraphics[width=0.3\textwidth]{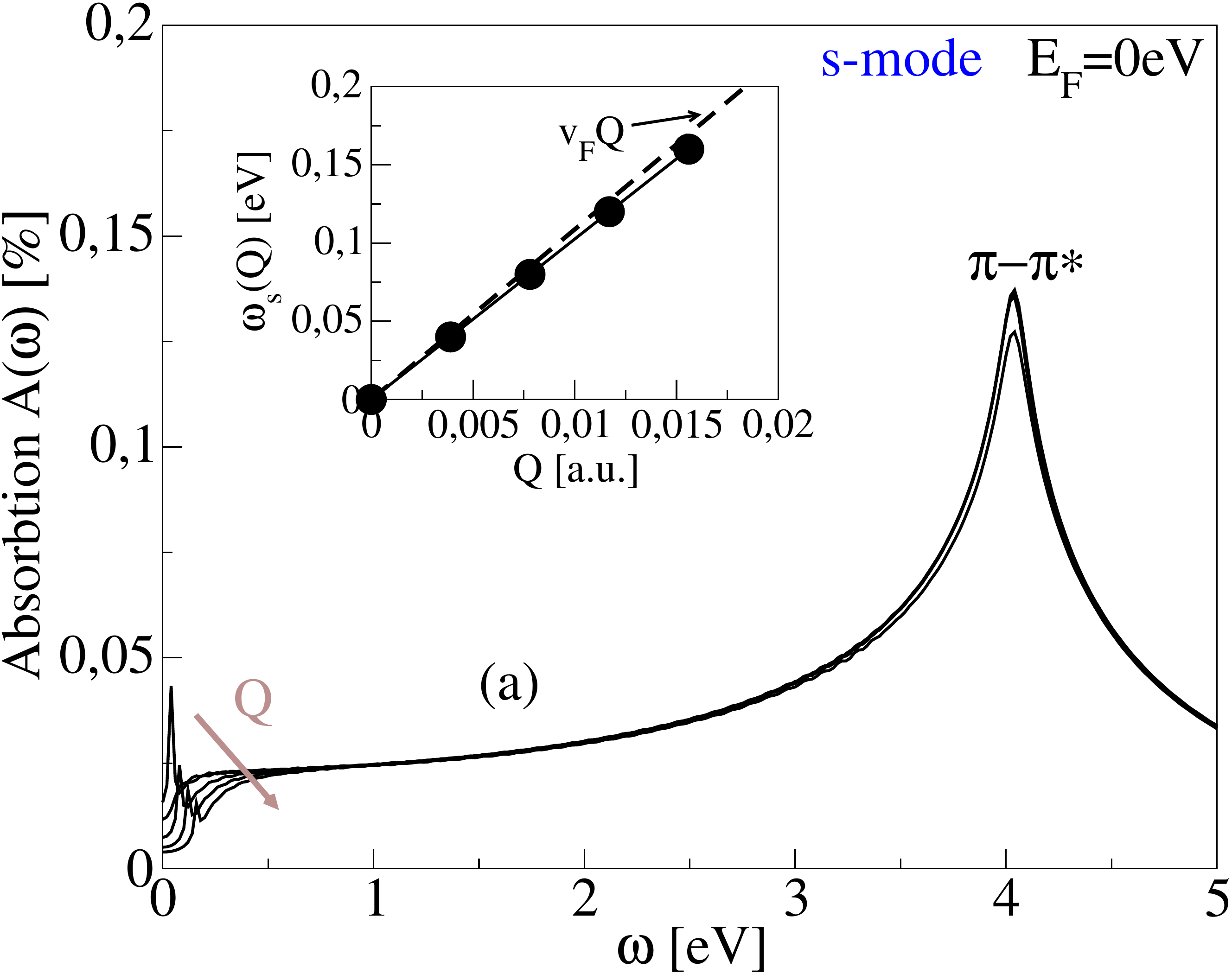}
\includegraphics[width=0.3\textwidth]{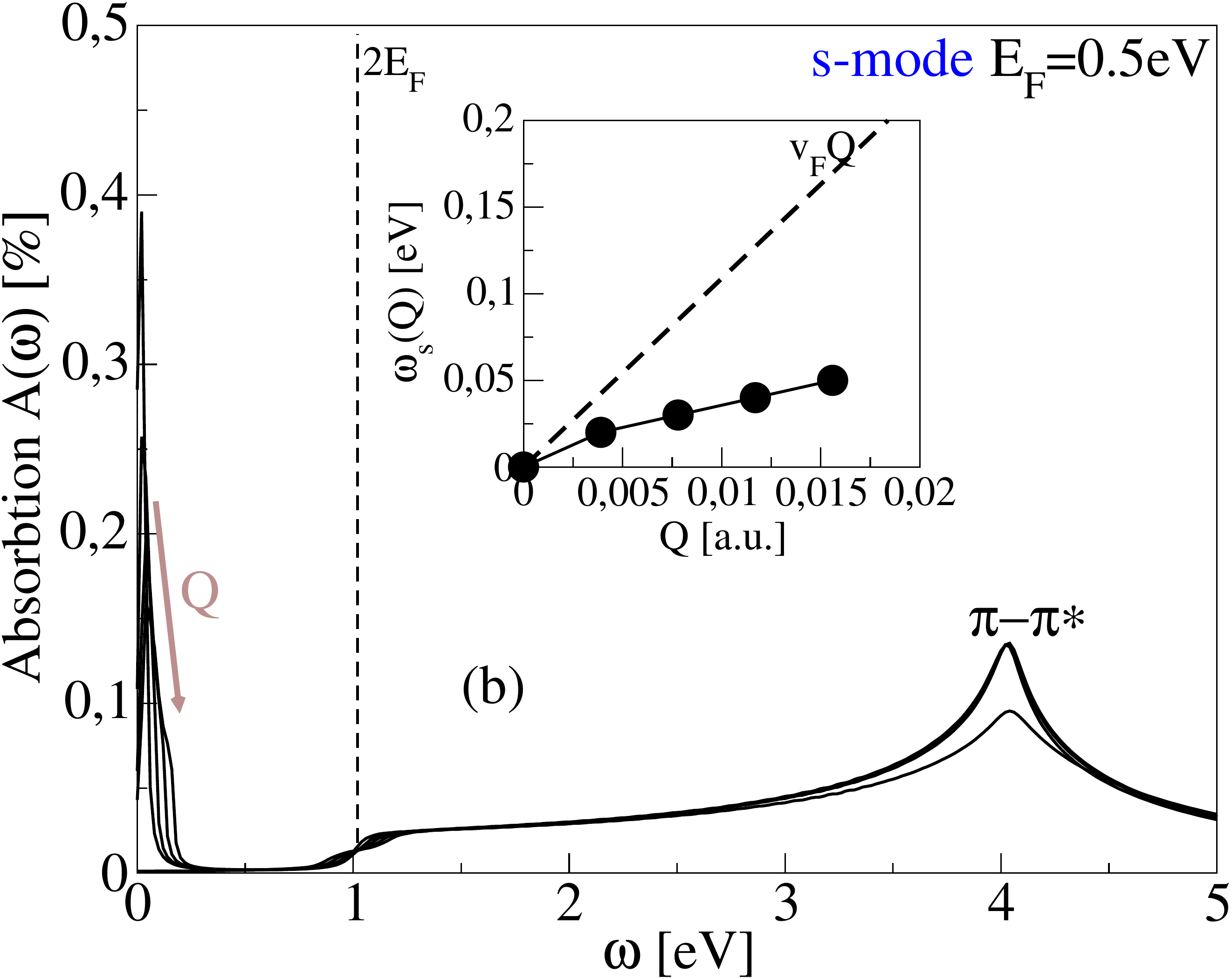}
\includegraphics[width=0.3\textwidth]{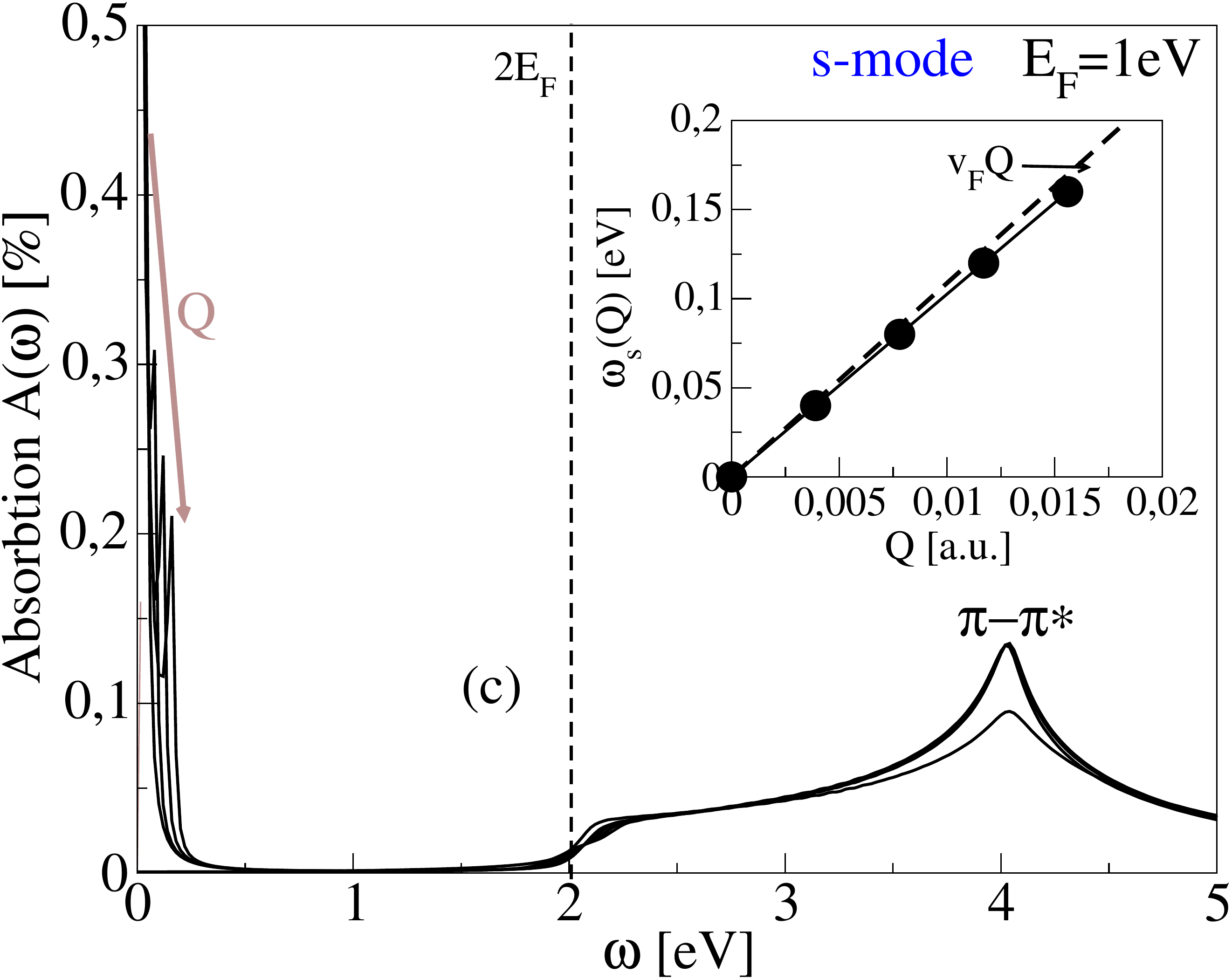}
\includegraphics[width=0.3\textwidth]{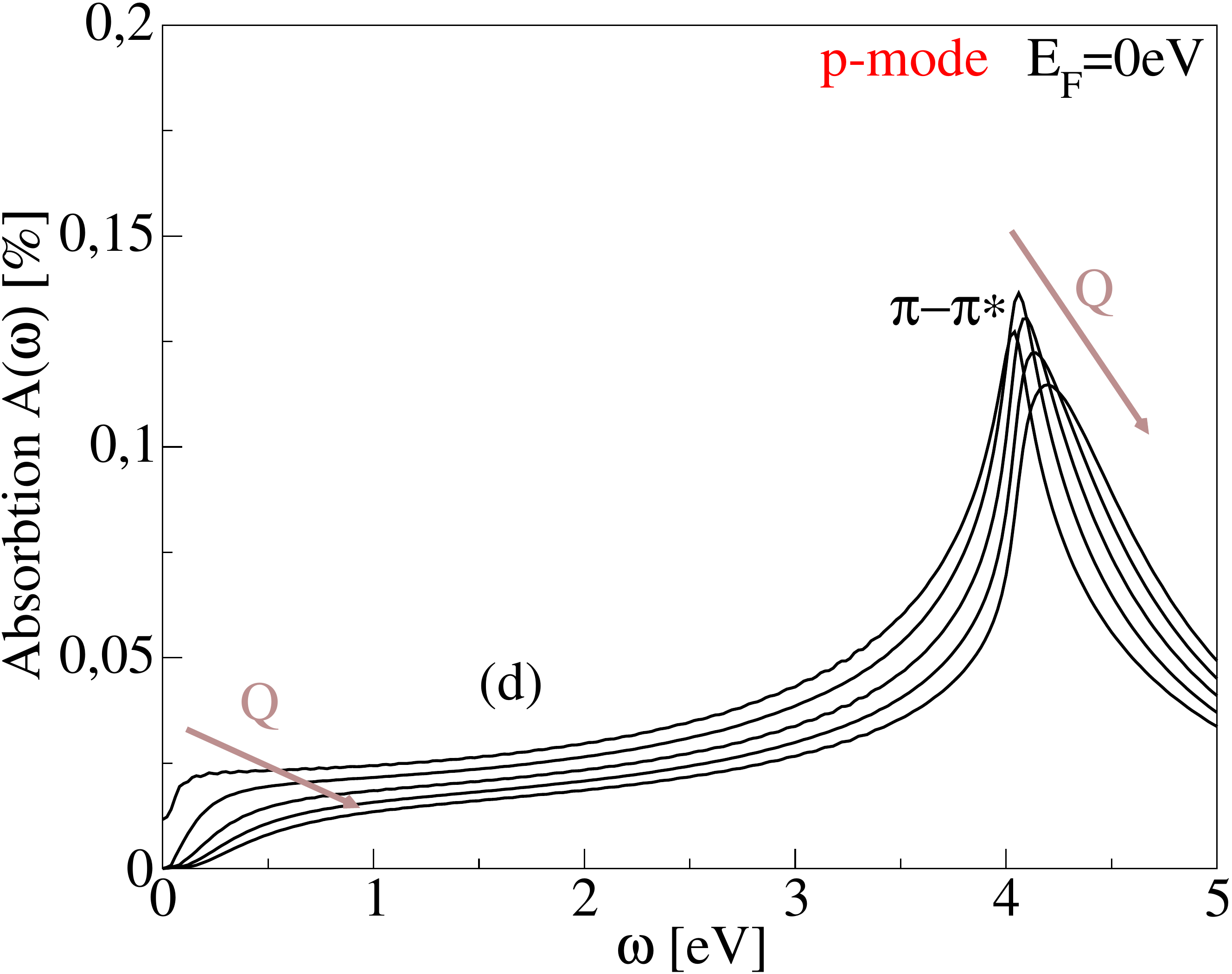}
\includegraphics[width=0.3\textwidth]{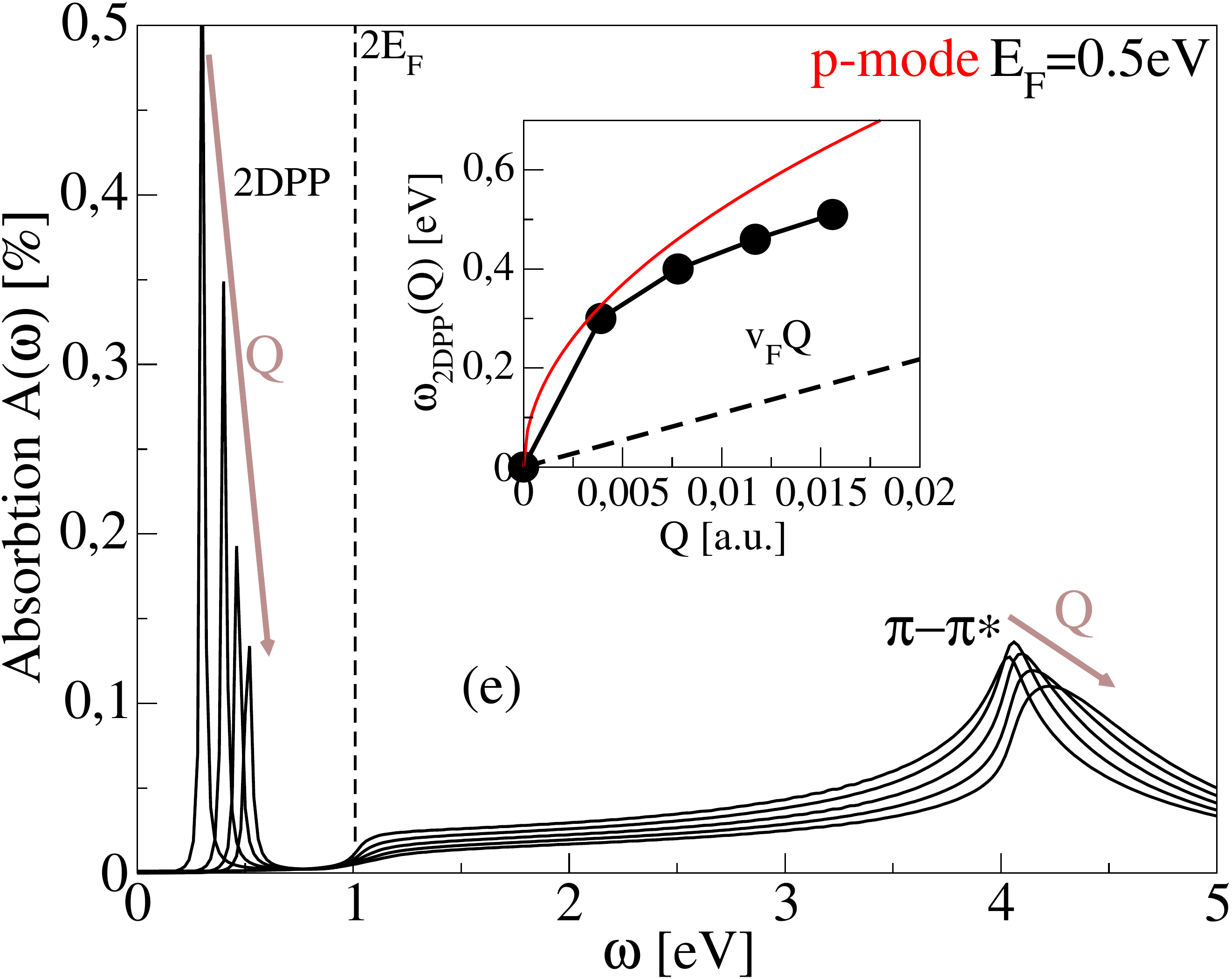}
\includegraphics[width=0.3\textwidth]{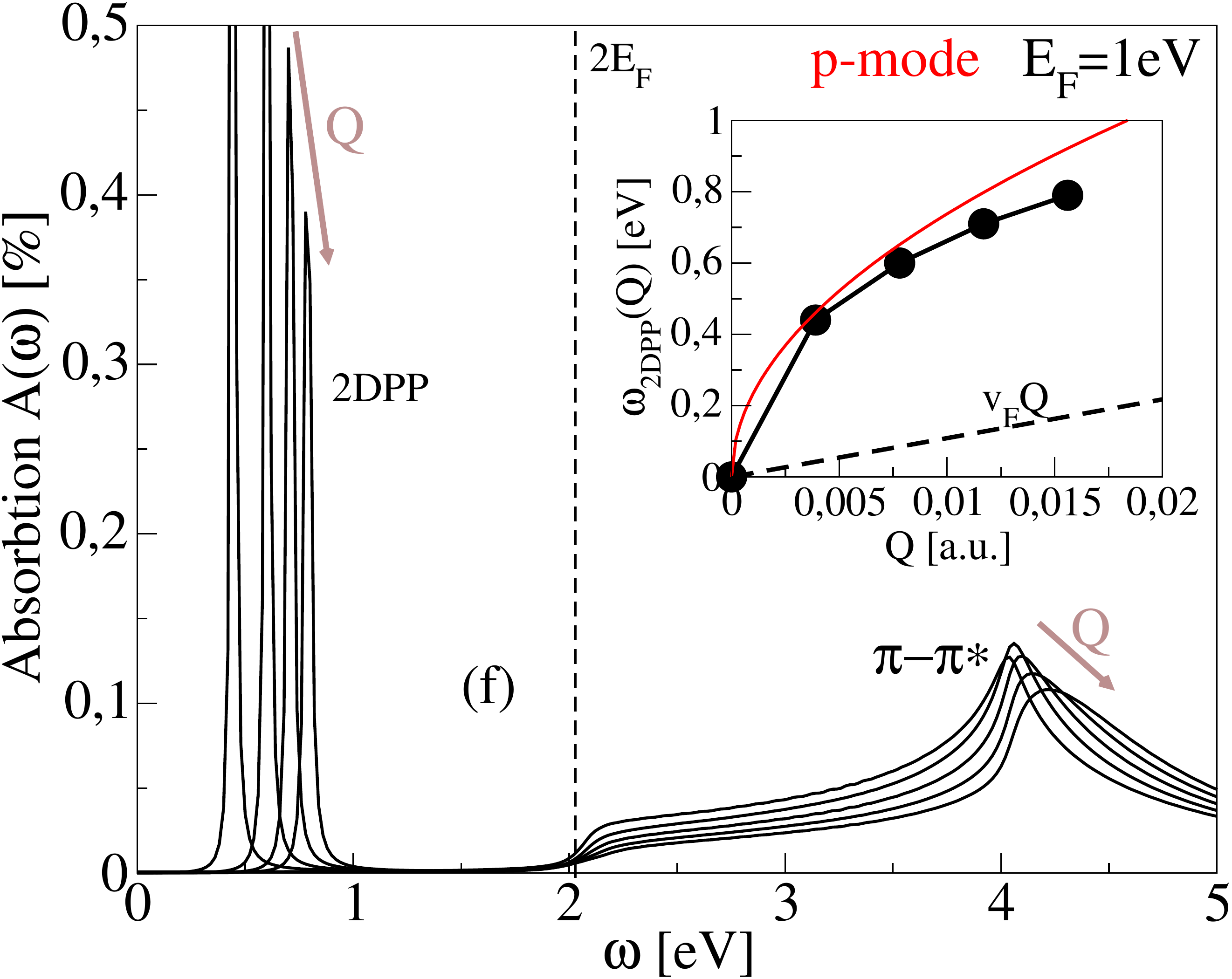}
\caption{(Color online) Optical absorption spectrum for $\mathbf{s}$ and $\mathbf{p}$ polarized incident electromagnetic field in graphene with different doping: (a,d) $E_F=0$, (b,e) $E_F=0.5\ \mathrm{eV}$ and (c,f) $E_F=1\ \mathrm{eV}$. In each of the graphs the five absorption spectrum are shown for five incident wavevectors along $\mathrm{\Gamma M}$ direction ${\bf Q}_n=n\Delta \mathrm{Q}\ {\bf y},\ \ n=0,1,2,3,4$, where $\Delta \mathrm{Q}=0.0039\ \mathrm{a.u.}$ The brown arrows indicate the direction of increasing wavevector. Insets in (a), (b) and (c) show the dispersion relations of low energy $\pi\rightarrow\pi$ intraband peaks, while insets in (e) and (f) show the dispersion relations of 2DPP. The dashed lines represent the upper edge ($v_FQ$) of the $\pi\rightarrow\pi$ intraband electron-hole continuum where the Fermi velocity is $v_F=\sqrt{3}ta/2$ and the hopping parameter is $t=2.7\ \mathrm{eV}$. Red solid lines represent the long wavelength approximation of the 2DPP dispersion relation $\sqrt{2E_F\mathrm{Q}}$.} 
\label{Fig4}
\end{figure*}

Figure \ref{Fig3}(a) shows the optical absorption coefficient (\ref{abscoef}) of $\mathrm{s}(x)$ or $\mathrm{p}(y)$ polarised electromagnetic fields in pristine graphene for normal incidence $\mathrm{Q}=0$. Optical absorption onset appears already at $\omega=0$ which is due to the gapless dipole active $\pi\rightarrow\pi^*$ interband transitions near K point of the BZ. In the infrared (IR) and visible regions ($\omega<3\ \mathrm{eV}$) absorption monotonically increases. The first absorption maximum, which appears in the ultraviolet (UV) region at $\omega=4\ \mathrm{eV}$, is a consequence of transitions between $\pi$ and $\pi^*$ bands along the $\mathrm{MM'}$ and $\mathrm{M\Gamma}$ directions of Q, as discussed in detail in \cite{Dino}. This resonance absorbs about $12\%$ of incident electromagnetic energy. In the far UV region $\omega>6\ \mathrm{eV}$ the spectrum shows more structures, which are due to optically active transitions between $\sigma$ and $\sigma^*$ bands, with the main peak at $\omega=13.9\ \mathrm{eV}$. This very strong excitation mode absorbs $30\%$ of the incident electromagnetic energy. 

Black solid line in the inset of Fig. \ref{Fig3}(a) shows the details of optical absorption in IR, visible and UV regions. In the whole IR region the absorption is close to the universal value of $\pi\alpha=2.293\%$ (denoted by the horizontal dashed line), as predicted experimentally in Refs. \cite{2.3posto1,2.3posto2}. In the far IR region ($\omega<200\ \mathrm{meV}$) the absorption begins to decrease faster until it reaches $A(\omega=0)$ value which is about half of the universal value $\pi\alpha$. However, the $A(\omega=0)$ value strongly depends on the damping constants $\eta_{\mathrm{intra}}$ and $\eta_{\mathrm{inter}}$ used in the calculation. Blue dashed line is the theoretical result taken from Ref. \cite{Louie}, and red dotted line is the experimental result taken from Ref. \cite{SSC}. We see that our absorption maximum is at a substantially lower energy (4.05 eV) then the 4.62 eV peak which appears in both, theoretical and experimental spectra. This is because the authors in Ref. \cite{Louie} in their GW-BSE calculation included quasiparticle corrections of the DFT band structure and also the exitonic effects, i.e. electron-hole interaction. Here the spectrum is calculated within the RPA which includes screening effects, although this screening is very inefficient in the optical limit ($\mathrm{Q\approx0}$), as discussed in Ref. \cite{Dino}. We see quite a nice agreement of our results with the theoretical result from Ref. \cite{Louie} in the whole IR and visible region, which will be the region of our main future interest. However, for $\omega<0.5\ \mathrm{eV}$ both theoretical results start disagreeing with the experimental spectrum which suddenly decreases below the universal value. This is probably due to the weak doping which causes a shift of optical absorption onset from $\omega=0$ to $\omega=2E_F$. Fig. \ref{Fig33} shows our theoretical results obtained using Eq. (\ref{eq3}) (yellow line) and Eq. (\ref{eq4}) (blue line) for the interband channel, which are then compared with the measured absorption spectrum for a graphene sample (red solid lines) taken from Ref. \cite{2.3posto2}. We see very good agreement in the frequency region $0.3\ \mathrm{eV}<\omega<1.2\ \mathrm{eV}$ (especially when using Eq. (\ref{eq4})), however we have slightly doped our graphene ($E_F=0.1\ \mathrm{eV}$) in order to achieve this result.

We want to emphasize here that $\Pi_{zz}({\bf Q}\approx 0,\omega)$ component of the full current-current response tensor is negligible in graphene \cite{RubioGraph}, as seen in Fig. \ref{Fig3}(b), so it is sufficient to use only $x$ and $y$ components for investigating electromagnetic response in graphene.

Now we show the results for the two electromagnetic modes in doped graphene, appearing within the window constrained by the intraband and interband continua: the usual TM mode (2D plasmon-polariton or 2DPP), shown for $\mathrm{Q\sim 10^{-7}}$ and $\mathrm{\omega\approx0}$ in Figs. \ref{Fig34}(a) and (b), and the TE mode \cite{TE1,TE2,tb2}, shown for $\mathrm{Q\sim 10^{-4}}$ and $\mathrm{\omega\sim E_F}$ in Figs. \ref{Fig35}(a) and (b). It can be seen that the 2DPP mode can be observed only if the incident EM wave is $\mathbf{p}$ polarised, while for the same $\mathrm{(Q,\omega)}$ values and $\mathbf{s}$ polarised EM wave the only feature appearing is the Drude peak intersected with the light line ($\mathrm{Qc}$). In the insets of Figs. \ref{Fig34}(a) and (b) we compare the dispersion of the 2DPP with the $\mathrm{\sqrt{Q}}$ 2D plasmon dispersion (when $\mathrm{c\rightarrow\infty}$). Here we see how coupling to the light waves influences the longitudinal plasmon mode and how its dispersion changes from $\mathrm{\sqrt{Q}}$ to $\mathrm{Qc}$ for very small $\mathrm{Q}$ and $\omega$ when the electronic excitations of graphene are coupled to EM waves \cite{Stern}. The effects of this coupling can also be seen in the second, TE, mode for $\mathrm{\omega\sim E_F}$. From Figs. \ref{Fig35}(a) and (b) we see that the energy peak of this mode is always just below $\mathrm{Qc}$, making it almost undistinguishable from the light line. Insets of these figures emphasise this even more.

In addition we present absorption intensities for $\mathrm{\omega\approx4\ eV}$ and $\mathbf{p}$  polarised EM wave, where the peak due to $\pi\rightarrow\pi^{\ast}$ transitions appears [Fig. \ref{Fig35}(c)]. It is clear from the figure that there is no coupling between these transitions and EM waves, which is an additional proof in the pristine graphene plasmon debate, that there is no so-called $\pi$ plasmon for $\mathrm{Q}\approx0$ while for the large Q wavevectors the plasmon is formed \cite{Dino,Nazarov}.

\subsubsection{$\mathrm{Q\gg Q_{light}}$}

Figures \ref{Fig4}(a)-(f) show absorption spectra for different dopings, incident polarizations and for five different wavevectors along the $\mathrm{\Gamma M}$ direction of the first Brillouin zone:
\[
{\bf Q}_n=n\Delta \mathrm{Q}\ {\bf \hat{y}},\ \ n=0,1,2,3,4, 
\]
where $\Delta \mathrm{Q}=0.0039\ \mathrm{a.u.}$. The brown arrows indicate direction of increasing wavevectors. It should be noted that for $n=0$ the whole frequency range in Fig. \ref{Fig4} corresponds 
to the radiative region, but for all nonzero wavevectors, $n>0$ (e.g. for $\mathrm{n}=1$ $\mathrm{Qc}=14.5\ \mathrm{eV}$), the whole frequency range corresponds to the evanescent region.

Figs. \ref{Fig4}(a), (b) and (c) show absorption spectra for $\mathrm{s}(x)$ polarized incident light for three different dopings $E_F=0$, $0.5$ and $1\ \mathrm{eV}$, respectively. For pristine graphene ($E_F=0$) far-IR absorption ($\omega<200\ \mathrm{meV}$) slowly decreases as $\mathrm{Q}$ increases and shows positively dispersive peaks, with the most intense one for $n=1$ when it reaches almost $5\%$ absorption. As Q increases the intensities of these peaks decrease. The black dots in the Fig. \ref{Fig4}(a) inset show the dispersion relation $\omega_s(\mathrm{Q})$ obtained by following the energies of these low energy peaks as functions of the wavevector $\mathrm{Q}$. We see that $\omega_s(\mathrm{Q})$ nicely follows the lower edge of the $\pi\rightarrow\pi^{\ast}$ interband continuum, ($v_F\mathrm{Q}$), for pristine graphene, shown by black dashed line. The interband $\pi\rightarrow\pi^{\ast}$ peak at $\omega\approx 4\ \mathrm{eV}$ remains dispersionless.

For doped graphene the $\pi\rightarrow\pi$ intraband excitation channel is open and absorption spectra for $\mathrm{s}(x)$ polarized light get strong maxima in the far-IR region, as shown in Figs. \ref{Fig4}(b) and (c). These absorption peaks are most intense for small $\mathrm{Q}$'s ($n=1,2$) and rapidly decrease as $\mathrm{Q}$ increases. The dispersion relations of this low energy peaks $\omega_s(\mathrm{Q})$ are shown in the inset of Figs. \ref{Fig4}(b) and (c). The upper edge of the intraband $\pi\rightarrow\pi$ electron-hole continuum is also shown (black dashed lines) for comparison. Around the Dirac cone it can be approximated by $v_F\mathrm{Q}$, where for the Fermi velocity we took $v_F=\sqrt{3}ta/2$ and where the hopping parameter of tight-binding model for graphene is $t\approx2.7\ \mathrm{eV}$ (as in \cite{Gr2013}). We can notice that both peaks are linearly dispersive, where for $E_F=1\ \mathrm{eV}$ $\omega_s(\mathrm{Q})$ follows exactly the $v_F\mathrm{Q}$ line. Because of the Pauli exclusion principle, as the doping increases the lower threshold for interband $\pi\rightarrow\pi^*$ transitions also increases. In other words, as these transitions are optically active the increased doping will open the optical gap and move the absorption onset towards higher energies. In Figs. \ref{Fig4}(b) and (c) we can notice this wide optical gap in the whole IR and visible region (depending on the doping) with the optical absorption onset at $2E_F$, denoted by the vertical dashed lines.

The absorption of $\mathrm{p}(y)$ polarised incident light is shown in Figs. \ref{Fig4}(d), (e) and (f). The parallel wave vector $\mathbf{Q}$ is chosen also to be in the $y$ ($\mathrm{\Gamma M}$) direction, so the response to this external electromagnetic field (for $\mathrm{Q}\ne0$) can be considered as longitudinal. It is evident from the results that this polarization gives generally more dispersive absorption spectrum in pristine graphene than the $\mathrm{s}(x)$ polarization. More specifically, we see that as $\mathrm{Q}$ increases the value of $A(\omega\approx 0)$ rapidly decreases, and the $\pi\rightarrow\pi^*$ peak ($\pi$ plasmon) becomes blue shifted and less intense. In doped graphene the Dirac cones are partially filled and Q2D plasma is formed. As already mentioned, this plasma supports longitudinal self-sustaining oscillations called 2D plasmons \cite{Gr2013}, or in the electrodynamic limit, 2DPP. These 2DPP have evanescent character, i.e. electrical fields which they produce in $z$ direction decay exponentially, as shown in Fig. \ref{Fig2}(b). This implies that these modes exist in the $\omega<\mathrm{Qc}$ region and cannot be excited (directly) by incident electromagnetic field which has fully radiative character. Figs. \ref{Fig4}(e) and (f) show absorption spectra of $\mathrm{p}(y)$ polarised incident field for doped graphene with $E_F=0.5$ and $1\ \mathrm{eV}$, respectively. For $\mathrm{Q}\geq \Delta \mathrm{Q}$ the incident field has evanescent character in the shown frequency range and becomes capable to excite 2DPP, which gives rise to the strong peaks in the optical gap $v_F\mathrm{Q}<\omega<2E_F$. Appearance of 2DPP causes strong screening which rapidly increases with $\mathrm{Q}$. One consequence of such screening is that for $\mathrm{p}(y)$ polarization there is no intraband $\pi\rightarrow\pi$ absorption maximum in the far-IR region (as was the case for $\mathrm{s}(x)$ polarization). The intensity of intraband $\pi\rightarrow\pi$ transitions is strongly reduced by the 2DPP screening field. The 2DPP dispersion relations $\omega_{\mathrm{2DPP}}(\mathrm{Q})$ shown by black dots in the inset of Figs. \ref{Fig4}(e) and (f), are compared with the simple long wavelength approximation $\sqrt{2E_F\mathrm{Q}}$ \cite{Gr2013} (red solid line). The apparent disagreement between these two dispersion relations for $\omega>0.8E_F$ is because in our calculations we considered both intraband and interband transitions, while to get the simplified dispersion one only accounts for the intraband transitions. In the region below $0.8E_F$ the agreement is better because there the interband contributions to 2DPP dispersion are negligible \cite{Kupcic,Kupcic1,Gr2013}.

\begin{figure}[t]
\centering
\includegraphics[width=\columnwidth]{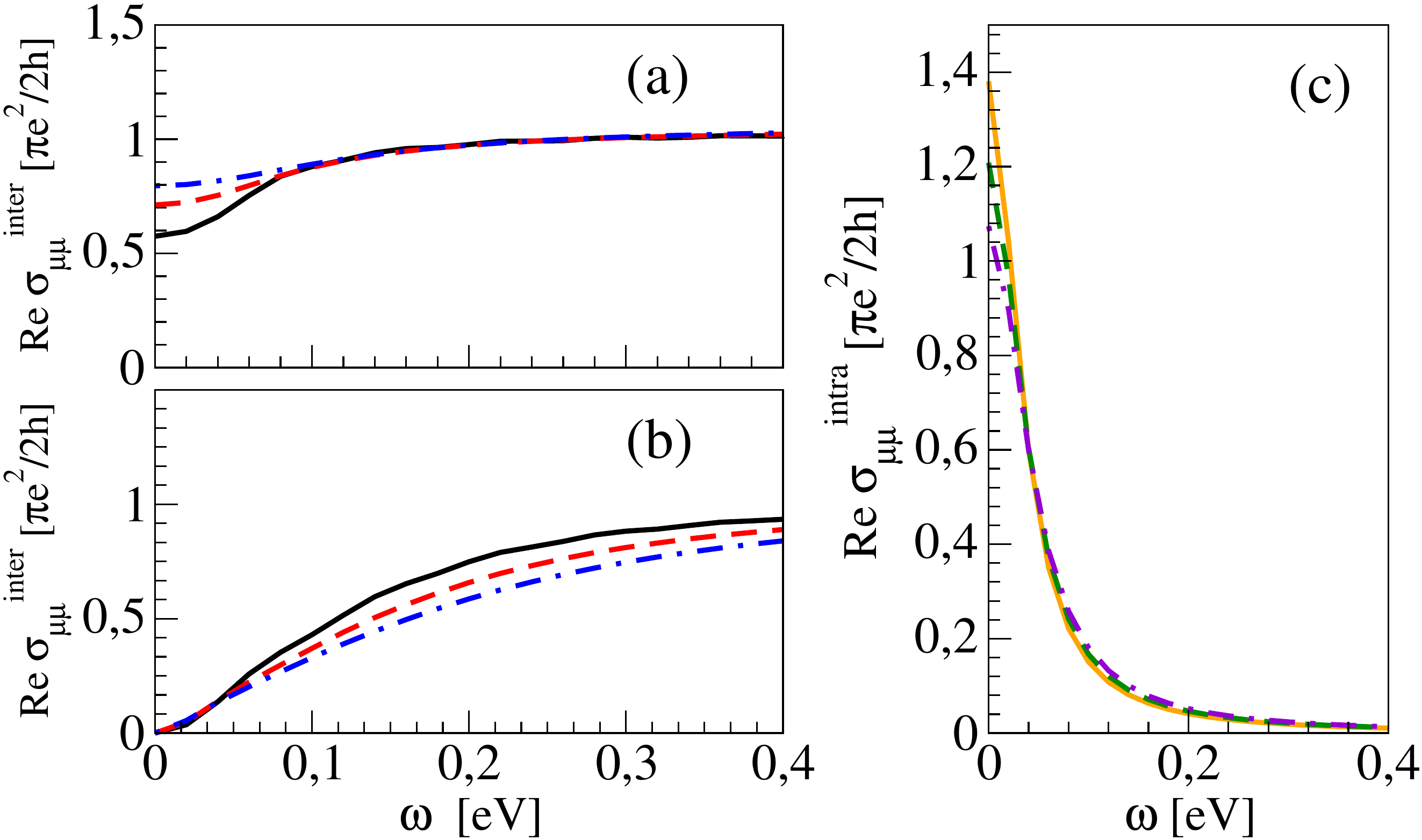}
\caption{(Color online) (a) Real part of interband optical conductivity in pristine graphene ($E_F=0$) calculated using Eq. (\ref{eq3}) where the damping constants are $\eta_{\mathrm{inter}}=60\ \mathrm{meV}$ (black solid), $\eta_{\mathrm{inter}}=90\ \mathrm{meV}$ (red dashed) and $\eta_{\mathrm{inter}}=120\ \mathrm{meV}$ (blue dashed-dotted). (b) Same as in (a) but using Eq. (\ref{eq4}). (c) Real part of intraband optical conductivity in pristine graphene ($E_F=0$) calculated using Eq. (\ref{eq2}) with $\eta_{\mathrm{intra}}=35\ \mathrm{meV}$ (yellow solid), $\eta_{\mathrm{intra}}=40\ \mathrm{meV}$ (green dashed) and $\eta_{\mathrm{intra}}=45\ \mathrm{meV}$ (purple dashed-dotted), and $T=300\ \mathrm{K}$ for all three cases. The wavevector is $\mathrm{Q=0}$ which corresponds to the electromagnetic field of normal incidence and polarization $\mu=x$ or $y$.}
\label{Fig5}
\end{figure}

\subsection{Optical conductivity}
In this section we analyse optical conductivity obtained from (\ref{sigmaexp}) in the system exposed to the homogeneous electrical field directed in the $\mu=x$ or $y$ direction and the wavevector is $\mathrm{Q}=0$. For $\mathrm{Q}=0$ the response is completely transverse, $\mathrm{s}(x)$ and $\mathrm{p}(y)$ polarizations are equivalent, the screening is inactive, so the screened ${\bf\Pi}$ and unscreened ${\bf\Pi}^0$ current-current response tensors (\ref{Pitotnew}) become equal. As the crystal local field effects in the $z$ direction are negligible, only the $G_z=G_z'=0$ component in $\Pi^0$ is nonzero. Also, because in the entire analysed frequency region $\frac{\omega L}{c}\rightarrow 0$, the form factor in (\ref{spectrfun}) becomes $F(G_z=0)\rightarrow 1$. Under these conditions the absorption (\ref{abscoef}) becomes proportional to optical conductivity (\ref{sigmaexp}), i.e. $A(\omega)\sim\sigma_{\mu\mu}(\omega)$, and therefore the discussion of the optical conductivity corresponds to the discussion of absorption spectra which were thoroughly analysed in Figs. \ref{Fig3} and \ref{Fig33} of the previous section. However, here we shall emphasize the Drude limit ($\omega\approx0$) and we focus on the case of pristine graphene only.

Fig. \ref{Fig5} shows the interband, Eqs. (\ref{eq3}) and (\ref{eq4}), and intraband, Eq. (\ref{eq2}), contributions to optical conductivity in pristine graphene. Our numerical method (being limited by finite $\mathbf{K}$ point sampling) is not especially appropriate to study different limits (e.g. $T\rightarrow0$, $\omega\rightarrow0$ and $\eta\rightarrow0$) when approaching the ballistic minimum conductivity, as explained in Refs. \cite{Ziegler1,Ziegler2,GraphRev}, but it can be very suitable to explore the Drude regime. We use the finite temperature ($T=300\ \mathrm{K}$) and energy damping ($\eta>0$) and in this case the intraband channel in pristine graphene is open [Fig. \ref{Fig5}(c)]. Interband contributions to optical conductivity Eqs. (\ref{eq3}) and (\ref{eq4}) give two different behaviours for $\omega\gtrsim0$ as seen in Fig. \ref{Fig5}(a) and (b). The first gives finite values up to $\omega=0$, while the second goes to 0 for $\omega=0$. In our case, where intraband channel presented in Fig. \ref{Fig5}(c) is open, it is easily seen that Eq. (\ref{eq4}) is more appropriate to use for the interband contribution to optical conductivity. Combining these two contributions one obtains good description of the experimental result from Ref. \cite{2.3posto2}.

\section{Conclusion}
\label{Conclu}
In this paper we presented a microscopic theory of electromagnetic response in Q2D layered crystals, in terms of dynamically screened current-current response tensor. In this approach the explicit knowledge of the electromagnetic field propagator can give us information about different polariton modes, both radiative and nonradiative, their spectra and intensities, their coupling to external fields and to other excitations in the crystal (e.g. phonons). Specifically it is straightforward to evaluate optical properties of such crystals (absorption, reflection, transmission and conductivity) and also to calculate higher-order many body processes. The key physical quantity is the current-current response tensor, calculated from first principles which implies inclusion of the realistic crystal structure, wave functions and electronic band structure. In order to test the developed formulation we calculated optical absorption and conductivity in a single layer graphene and compared with recent experimental and theoretical results. The obtained results agree well with the measurements and experiments in IR and visible frequency regions, though the use of RPA is not capable to give correct excitonic effect. The theory is therefore suitable for the study of the layered nanostructures in the IR frequency range which is nowadays of great importance in plasmonics, which is a growing branch of theoretical but also of applied physics.

\begin{acknowledgements}
The authors are grateful to Donostia International Physics Center (DIPC) and Pedro M. Echenique for hospitality and financial support during various stages of this work. We also thank I. Kup\v ci\'c and V. M. Silkin for useful discussions. Computational resources were provided by the DIPC computing center.
\end{acknowledgements}


\begin{thebibliography}{99}
\bibitem{comment} In principle, whether we use the charge-charge or current-current response function for calculating the polarizability of the system we should end up with the same result 
(as required by the charge conservation and gauge invariance \cite{Schrieffer}), but in the numerical calculation this is not so easy to achieve due to the finite k point sampling and the 
use of finite broadening parameter.
\end{thebibliography}
\end{document}